\documentclass[A4,aps,twocolumn]{revtex4}
\usepackage{amsmath,lscape,sidecap,epsfig}
\usepackage{booktabs}
\usepackage{color}

\def\beq{\begin{equation}}
\def\eeq{\end{equation}}
\def\beqa{\begin{eqnarray}}
\def\eeqa{\end{eqnarray}}
\def\ban{\begin{eqnarray*}}
\def\ean{\end{eqnarray*}}
\def\bi{\begin{itemize}}
\def\ei{\end{itemize}}

\usepackage{ulem}

\newcommand{\unit}[1]{\ensuremath{\, \mathrm{#1}}}
\newcommand{\pr}[1]{\ensuremath{\left[#1\right]}}
\newcommand{\pc}[1]{\ensuremath{\left(#1\right)}}

\begin{document}

\title{Description of light clusters in relativistic nuclear models } 

\author{M\'arcio Ferreira}
\affiliation{Centro de F\'{i}sica Computacional, Department of Physics,
University of Coimbra, P-3004-516 Coimbra, Portugal}
\author{Constan\c{c}a Provid\^encia}
\affiliation{Centro de F\'{i}sica Computacional, Department of Physics,
University of Coimbra, P-3004-516 Coimbra, Portugal}

\begin{abstract}
Light clusters are included in the equation of state of nuclear matter within
the relativistic mean field theory. The effect of the cluster-meson coupling
constants on the dissolution density is discussed. Theoretical and
experimental constraints are used to fix the cluster-meson couplings.
 The relative light cluster
fractions are calculated for
asymmetric matter in chemical equilibrium at finite temperature. It is found
that above  $T=5$ MeV  deuterons and tritons are the clusters in larger
abundances. The results do not depend strongly on the relativistic mean field
interaction chosen. 

\end{abstract}
\maketitle

\vspace{0.50cm}
PACS number(s): {21.65.+f, 24.10.Jv, 26.60.+c, 95.30.Tg}
\vspace{0.50cm}

\section{Introduction}
To gain a deeper understanding of the physics involved in stellar core collapse, 
supernova explosion and protoneutron star evolution, an equation of state capable
of describing matter ranging from very low densities to few times the saturation 
density and from zero temperature to a few $\unit{MeV}$ is needed.

The
  crust of the star is essentially determined  by the low density region
  of the equation of state (EOS), while the high density part will be important to define properties such as the star mass and radius.
Understanding the  crust constitution of a neutron star is an important issue 
because it influences the cooling process
of the star and  plays a decisive role  on  quantities such as the neutrino emissivity
and gravitational waves emission.

In the outer crust,  where the density is lower, the formation of light clusters in 
nuclear matter will be energetically favorable at finite temperature, as
 in core-collapse supernovae or neutron star mergers, whereas in catalyzed
 cold neutron stars only heavy nuclei appear.
 At very low densities
and moderate temperatures, the few body correlations are expected to become important and 
the system minimizes its free energy by forming light nuclei like
deuterons  ($d\equiv\, ^{2}\text{H}$), tritons ($t\equiv\, ^3\text{H}$), helions
($h\equiv\, ^3\text{He}$) and $\alpha$-particles ($^4\text{He}$)
 due to the increased entropy \cite{ropke2009,typel2010}. Eventually, these clusters 
will dissolve at higher densities due to Pauli blocking resulting in an
homogeneous matter \cite{ropke2009}.
In particular, the cooling process of a protoneutron star is affected by the
appearance of light clusters \cite{light3}.

The inclusion of light clusters ($d$, $h$, $t$ and $\alpha$ particles) in the nuclear matter EOS is discussed in \cite{typel2010}, 
where the most important thermodynamical quantities are calculated within a density-dependent relativistic model.
The conditions for the liquid-gas phase transition are obtained, and it is seen how the binodal section
is affected by the inclusion of these clusters. Moreover, an EOS is obtained starting at low densities 
with clusterized matter up to high density cluster-free homogeneous matter.  In this work the density and
temperature dependence of the in-medium binding energy of the clusters was determined
within a quantum statistical approach \cite{ropke} and included in a phenomenological way
in the relativistic mean field (RMF) model with density dependent couplings.

The $\alpha$ particle is the most strongly bound system among all light nuclei and it certainly plays a role in 
nuclear matter as has been pointed out in \cite{ls91,shen,ropke,hor06,typel2010}.
Lattimer and Swesty \cite{ls91} worked out the EOS in the compressible extended liquid 
drop model based on a non-relativistic framework  appropriate for supernova simulations, for a wide range of densities, proton fractions and temperatures, including the contribution of
$\alpha$ particle clusters. An excluded volume prescription is used to model the dissolution of 
$\alpha$ particles at high densities. The same is done by H. Shen et al. in \cite{shen} where non-uniform 
matter composed of protons, neutrons, $\alpha$ particles and a single species of heavy nuclei is 
described with the Thomas-Fermi approximation and the TM1 parametrization of the non-linear Walecka 
model (NLWM). At low densities, these particles are described by classical gases.

In \cite{hor06} the virial expansion of low-density nuclear matter composed by neutrons, protons, and $\alpha$ particles is 
presented, and it is shown that the predicted $\alpha$ particles concentrations differ from the predictions 
of the EOS proposed in \cite{ls91} and \cite{shen}. The virial expansion was
extended to include other light clusters with $A\le 4$
\cite{oconnor2007,light4}. All possible light clusters were also included in
recent EOS appropriate for  simulations of core-collapse supernovae based in
a nuclear statistical equilibrium model, including excluded volume effects
for nuclei \cite{hempel2010}.

The aim of the present work is to study  the dissolution density  of light
clusters, i.e. the
density above which light clusters do not exist anymore,  in uniform nuclear
matter within the framework of the NLWM \cite{bb}. The dependence of the dissolution density
on the cluster-meson couplings as well as on the proton-fraction will be
studied. In many models which include light clusters the dissolution density
is determined by excluded volume effects \cite{ls91,shen,hempel2010}. In
Ref. \cite{light4}, a statistical excluded volume model was compared with two
quantum many-body models, a generalized relativistic mean-field model \cite{typel2010} and
a quantum statistical model \cite{ropke2009,ropke2011}.  It was shown that the
excluded volume description works reasonably well at high
temperatures, partly due to a reduced Pauli blocking. However, at low temperatures the excluded volume approach shows
crucial deviations from the statistical approach.

 In \cite{alphas}, using a
RMF approach,  the
coupling of the $\alpha$-clusters to the $\omega$-meson was employed to describe the
dissolution. In this reference it was assumed that the intensity of the $\omega$
meson-cluster coupling was proportional to the mass number of the light
cluster and no coupling to the $\sigma$-meson was included.

  The light clusters ($d$, 
$h$, $t$ and $\alpha$ particles) are treated as point like particles (without internal 
structure) within the RMF approximation. Just as  the nucleons in nuclear matter,
the clusters interact through  the exchange of  $\sigma$, $\omega$ and $\rho$
meson fields. In order to understand how dependent are the results on the NLWM
parametrization we consider  three different parametrizations; NL3 \cite{nl3}, with a quite large symmetry energy and
incompressibility at saturation and which was fitted in order to
reproduce the ground state properties of both stable and unstable
nuclei; FSU \cite{fsu} and IU-FSU \cite{iufsu}, which were
accurately calibrated to simultaneously describe the GMR in
$^{90}$Zr and $^{208}$Pb, and the IVGDR in $^{208}$Pb and still
reproduce ground-state observables of stable and unstable nuclei.
 Furthermore, the IU-FSU was also refined to describe neutron stars with high
 masses ($\sim2M_\odot$). 

In Ref. \cite{typel2010}, where a generalized RMF model with density dependent
couplings was developed including clusters as explicit
degrees of freedom, the  medium effects on the binding energy of the
clusters were introduced explicitly through density and temperature dependent terms.
We aim at getting a description of the light clusters in nuclear matter that
is simpler to implement, yet more realistic than the excluded volume
mechanism.

A brief review of the formalism is presented in section \ref{formalism}, the
results and discussion of the meson-cluster coupling constants are given in
section \ref{results} and conclusions are drawn in section \ref{conclusions}.

\section{The Formalism}\label{formalism}

We consider a system of protons and neutrons
interacting with and through an isoscalar-scalar field $\sigma$ with mass
$m_s$,  an isoscalar-vector field $\omega^{\mu}$ with mass $m_v$, an 
isovector-vector field  $\mathbf b^{\mu}$ with mass
$m_\rho$ and light clusters: $d$, $h$, $t$, and $\alpha$ particles.
%
The Lagrangian density of this system reads:
\begin{equation*}
\mathcal{L}=\sum_{j=p,n,t,h}\mathcal{L}_{j}+\mathcal{L}_{{\alpha }}+
\mathcal{L}_d+ \mathcal{\,L}_{{\sigma }}+ \mathcal{L}_{{\omega }} + 
\mathcal{L}_{{\rho }} + \mathcal{L}_{\omega \rho},
\label{lag}
\end{equation*}
where
the Lagrangian $\mathcal{L}_{j}$ is
\begin{equation}
\mathcal{L}_{j}=\bar{\psi}_{j}\left[ \gamma _{\mu }iD^{\mu }_j-M^{*}_j\right]
\psi _{j}  \label{lagnucl},
\end{equation}
with 
\begin{eqnarray}
iD^{\mu }_j &=&i\partial ^{\mu }-g_v^j\omega^{\mu }-\frac{g_{\rho }^j}{2}{\boldsymbol{\tau}}%
\cdot \mathbf{b}^{\mu } 
, \label{Dmu} \\
M^{*}_j &=&M_j-g_{s}^j\sigma, \quad j=p,n,t,h
\label{Mstar}
\end{eqnarray}
where $M=(M_p+M_n)/2=938.918695\unit{MeV}$, $M_{h}=3M - B_{h}$, $M_{t}=3M - B_{t}$, $M_{\alpha}= 4 M - B_{\alpha}, \,\,\, M_d= 2 M - B_d$,  with the binding 
energies given in Table \ref{tab:Typel}. 
The $\alpha$ particles and the deuterons are described as in \cite{typel2010}
\begin{eqnarray}
\mathcal{L}_{\alpha }&=&\frac{1}{2} (i D^{\mu}_{\alpha} \phi_{\alpha})^*
(i D_{\mu \alpha} \phi_{\alpha})-\frac{1}{2}\phi_{\alpha}^* \pc{M_{\alpha}^*}^2
\phi_{\alpha},\\
\mathcal{L}_{d}&=&\frac{1}{4} (i D^{\mu}_{d} \phi^{\nu}_{d}-
i D^{\nu}_{d} \phi^{\mu}_{d})^*
(i D_{d\mu} \phi_{d\nu}-i D_{d\nu} \phi_{d\mu})\nonumber\\
&&-\frac{1}{2}\phi^{\mu *}_{d} \pc{M_{d}^*}^2 \phi_{d\mu},
\end{eqnarray}
where
\begin{equation}
iD^{\mu }_j = i \partial ^{\mu }-g_v^j \omega^{\mu }
\end{equation}
and
\begin{equation}
 M_{j}^*=M_{j}-g_s^j\sigma
\label{mboson}
\end{equation}
with $j=\alpha,d$. The meson Lagrangian densities are 
\begin{eqnarray*}
\mathcal{L}_{{\sigma }} &=&\frac{1}{2}\left( \partial _{\mu }\sigma \partial %
^{\mu }\sigma -m_{s}^{2}\sigma ^{2}-\frac{1}{3}\kappa g_s^3 \sigma ^{3}-\frac{1}{12}%
\lambda g_s^4\sigma ^{4}\right)  \\
\mathcal{L}_{{\omega }} &=&\frac{1}{2} \left(-\frac{1}{2} \Omega _{\mu \nu }
\Omega ^{\mu \nu }+ m_{v}^{2}\omega_{\mu }\omega^{\mu }
+\frac{1}{12}\xi g_{v}^{4}(\omega_{\mu}\omega^{\mu })^{2} 
\right) \\
\mathcal{L}_{{\rho }} &=&\frac{1}{2} \left(-\frac{1}{2}
\mathbf{B}_{\mu \nu }\cdot \mathbf{B}^{\mu
\nu }+ m_{\rho }^{2}\mathbf{b}_{\mu }\cdot \mathbf{b}^{\mu } \right)\\
\mathcal{L}_{\omega \rho } &=& \Lambda_v g_v^2 g_\rho^2 \omega_{\mu }\omega^{\mu }
\mathbf{b}_{\mu }\cdot \mathbf{b}^{\mu }
\end{eqnarray*}
where $\Omega _{\mu \nu }=\partial _{\mu }\omega_{\nu }-\partial _{\nu }\omega_{\mu }$, 
$\mathbf{B}_{\mu \nu }=\partial _{\mu }\mathbf{b}_{\nu }-\partial _{\nu }
\mathbf{b}
_{\mu }-g_{\rho }(\mathbf{b}_{\mu }\times \mathbf{b}_{\nu })$. 
The  parameters of the models  are: the masses and
the coupling parameters $g_s$, $g_v$, $g_{\rho}$ of the mesons and $\lambda$, 
$\kappa$, $\xi$, and  $\Lambda_v$ of the nonlinear meson terms.
In the above Lagrangian density $\boldsymbol {\tau}$ is
the isospin operator. When the NL3 parametrization is used, $\xi$ and $\Lambda_v$ are set
equal to zero. The FSU and IU-FSU parametrizations include the nonlinear $\omega\rho$ term.

The equations of motion in the relativistic mean field approximation
 for the meson fields are given by 
\begin{align*}
 &m_s^2\sigma +\frac{\kappa}{2}g_s^3\sigma^2+\frac{\lambda}{6}g_s^4\sigma^3=\sum_{i=p,n,t,h}g_s^i\rho_s^i+\sum_{i=d,\alpha}g_s^i\rho_i\\
&m_v^2\omega^0+\frac{1}{6}\xi g_v^4\pc{\omega^0}^3+2\Lambda_v g_v^2g_\rho^2\omega^0\pc{b_3^{0}}^2=\sum_{i=p,n,t,h,d,\alpha}g_v^i\rho_i\\
&m_\rho^2b_3^{0}+2\Lambda_v g_v^2g_\rho^2\pc{\omega^0}^2b_3^{0}=\sum_{i=p,n,t,h}g_\rho^i I_3^i\rho_i 
\end{align*}
where, $\rho_s^i$ is the scalar density given by
\begin{equation*}
  \rho_s^i=\frac{2S^i+1}{2\pi^2}\int_0^{k_f^i}k^2dk\frac{M_i-g_s^i\sigma}{\sqrt{k^2+\pc{M_i-g_s^i\sigma}^2}},
\end{equation*}
$\rho_i$ is the vector density,  $I_3^i$ ($S^i$) is the isospin (spin) of the specie $i$ for $i=n,p,t,h$. At zero temperature all the $\alpha$ and deuteron 
populations will condense in a state of zero momentum, therefore $\rho_\alpha$ and $\rho_d$ correspond to
the condensate density of each specie.

The values of the parameters for each one of the relativistic effective interactions, discussed below, are in Table \ref{parameter_sets} and the 
corresponding properties of infinite nuclear matter at saturation density are shown in Table \ref{parameter_properties}.
\begin{table}[h]
\begin{center}
\caption{Parameters for the cluster binding energy shifts defined in
  Eqs. (\ref{bind1})- (\ref{bind3}) at $T=0$ MeV (taken from \cite{typel2010}).}
\begin{tabular}{ccccc}
\hline
\hline
Cluster $i$ & $a_{i,1}$ & $a_{i,2}$ &$a_{i,3}$& $B_i^0$ \cite{audi} \\
 & $\pc{\unit{MeV}^{5/2}\unit{fm}^3}$ & $\pc{\unit{MeV}}$ & $\pc{\unit{MeV}}$ \\
\hline
$t$ & $69516.2$ & $7.49232$& $-$ & $8.481798$  \\
$h$ & $58442.5$ & $6.07718$& $-$ & $7.718043$  \\
$\alpha$ & $164371$ & $10.6701$& $-$ & $28.29566$  \\
$d$ & $38386.4$ & $22.5204$& $0.2223$ & $2.224566$  \\
\hline
\hline
\end{tabular}
\label{tab:Typel}
\end{center}
\end{table}

\begin{table}[h]
\begin{center}
\caption{Parameter sets for the three models used in this work.}
\begin{tabular}{cccc}
\hline
\hline
 & NL3 \cite{nl3}& FSU \cite{fsu} & IU-FSU \cite{iufsu}  \\
\hline
$m_s\,\pc{\unit{MeV}}$ & $508.194 $ & $491.500$ & $491.500$  \\ 
$m_v\,\pc{\unit{MeV}}$ & $782.501$ & $782.500$ & $782.500$  \\ 
$m_\rho\,\pc{\unit{MeV}}$ & $763.000$ & $763.000$ & $763.000$  \\ 
$g_s^2$ & $104.3871$ & $112.1996$ & $99.4266$  \\ 
$g_v^2$ & $165.5854$ & $205.5469$ & $169.8349$  \\ 
$g_\rho^2$ & $79.6000$ & $138.4701$ & $184.6877$  \\ 
$\kappa\,\pc{\unit{MeV}}$ & $3.8599$ & $1.4203$ & $3.3808$  \\ 
$\lambda$ & $-0.015905 $ & $+0.023762 $ & $+0.000296$  \\ 
$\xi$ & $0.00 $ & $0.06 $ & $0.03 $  \\ 
$\Lambda_v$ & $ 0.000$ & $0.030 $ & $0.046$  \\
\hline
\hline
\end{tabular}
\label{parameter_sets}
\end{center}
\end{table}

\begin{table}[h]
\begin{center}
\caption{Bulk parameters characterizing the behavior of infinite symmetric
  nuclear matter at saturation density: the saturation density $\rho_0$,
the  binding energy $E/A$, the incompressibility $K$, the symmetry energy $a_{sym}$ and
  its slope $L$, and the nucleon effective mass $M^*$.}
\begin{tabular}{cccc}
\hline
\hline
 & NL3 & FSU  & IU-FSU   \\
\hline
$\rho_0\pc{\unit{fm}^{-3}}$ & $0.148$ & $0.148$ & $0.155$  \\ 
$E/A\pc{\unit{MeV}}$ & $-16.24$ & $-16.30$ & $-16.40$  \\ 
$K\pc{\unit{MeV}}$ & $271.5$ & $230.0$ & $231.2$  \\ 
$a_{sym}\pc{\unit{MeV}}$ & $37.29$ & $32.59$ & $31.30$  \\ 
$L\pc{\unit{MeV}}$ & $118.2$ & $60.5$ & $47.2$  \\ 
$M^*/M$ & $0.60$ & $0.62$ & $0.62$  \\ 
\hline
\hline
\end{tabular}
\label{parameter_properties}
\end{center}
\end{table}

The total energy density is given by
\begin{align*}
{\cal E}=&\sum_{i=p,n,t,h}\pc{\frac{2S^i+1}{2\pi^2}\int_0^{k_f^i}k_i^2dk_i\sqrt{k_i^2+(M_i-g_s^i\sigma)^2}}\\
&+\sum_{i=p,n,t,h}\pc{g_v^i\omega^0\rho_i+g_\rho^i b_3^{0}I_3^i\rho_i}+\frac{1}{2}m_s^2\sigma^2+\frac{1}{6}\kappa g_s^3\sigma^3\\
&+\frac{1}{24}\lambda g_s^4\sigma^4 -\frac{1}{2}m_v^2 \omega_0 \omega^0-\frac{1}{24}\xi g_v^4\pc{\omega^0}^4 -\frac{1}{2}m_\rho^2\pc{b_3^{0}}^2\\
&-\Lambda_v g_v^2g_\rho^2\omega_0 \omega^0\pc{b_3^{0}}^2+\sum_{i=\alpha,d}\pc{M_i-g_s^i\sigma +g_v^i\omega^0}\rho_i.\label{eq:energy_density}
\end{align*} 
At zero temperature the free energy density $ {\cal F}$  and the energy density coincide.

We define the global proton fraction $Y_{p}$ as
\begin{align*}
 Y_{p}&=\frac{\rho_p}{\rho}+2\frac{\rho_\alpha}{\rho}+\frac{\rho_d}{\rho}+2\frac{\rho_h}{\rho}+\frac{\rho_t}{\rho}.
\end{align*}
where $\rho=\rho_p+\rho_n+4\rho_\alpha+2\rho_d+3\rho_h+3\rho_t$ is the total baryonic density.
Defining the mass fraction $y_i=A_i(\rho_i/\rho)$ of the various species we obtain 
$$
Y_{p}=y_{p}+\frac{1}{2}y_{\alpha}+\frac{1}{2}y_{d}+\frac{2}{3}y_{h}+\frac{1}{3}y_{t}
$$
for the global proton fraction.
\subsection{Cluster binding energy}
In \cite{typel2010} using  a quantum statistical approach, the following empirical 
quadratic form was proposed for the medium-dependent binding energy shift at zero temperature 
of the various clusters
\begin{equation}
 \Delta B_i(\rho,T=0)=-\tilde{\rho}_i\pr{\delta B_i(0)+\frac{\tilde{\rho}_i}{2B_i^0}\delta B_i^2(0)}
\label{bind1}
\end{equation}  
where
\begin{equation*}
\tilde{\rho}_i=\frac{2}{A_i}\pr{Z_i \rho_p^{tot}+N_i \rho_n^{tot}}.
\end{equation*}  
The total nucleon density is given by $\rho=\rho_p^{tot}+\rho_n^{tot}$ and
\begin{equation}
\delta B_i(T=0)=\frac{a_{i,1}}{a_{i,2}^{3/2}}
\label{bind2}
\end{equation}
for $i=t,h,\alpha$, and 
\begin{equation}
\delta B_d(T=0)=\frac{a_{d,1}}{2a_{d,3}^2 a_{d,2}^{3/2}}
\label{bind3}
\end{equation}
for the deuteron.
The parameters $a_{i,1}$, $a_{i,2}$  and $a_{d,3}$, taken from \cite{typel2010}, are listed for symmetrical nuclear matter ($Y_p=0.5$) 
in Table \ref{tab:Typel}.

The density where a cluster becomes unbound, $\rho_{dis}$, referred as dissolution density,
 is given by 
\begin{equation}
\label{eq:typel}
 B_i(\rho_{dis})=B_i^0+\Delta B_i=0.
\end{equation}
The dissolution densities obtained for the light clusters considered at zero temperature and in 
symmetric nuclear matter are reported in the Table \ref{tab:resul_energ}.  

\begin{table}[h]
\begin{center}
\caption{Densities at which the clusters become unbound given by equation
  (\ref{eq:typel}) for $T=0$ MeV. }
\begin{tabular}{ccccc}
\hline
\hline
 & $t$ & $h$ & $\alpha$ & $d$ \\
\hline
$\rho_{dis}\,\pc{\unit{fm}^{-3}}$ & $0.00183175$ & $ 0.00144835$ & $0.00439226$ & $0.00044815$   \\  							      
\hline
\hline
\end{tabular}
\label{tab:resul_energ}
\end{center}
\end{table}
\section{Results and discussions} \label{results}

Our results will be reported in the present section. The
calculations will all be performed at zero temperature except for
Section \ref{results4} where results at finite temperature will also
be presented.

In the following we will consider each type of light
clusters separately in order to study the dissolution density of each
of them in nuclear matter.

We assume for the cluster couplings the form
\begin{align}\label{eq:cc}
 g_{v}^j&=\eta_j g_v \nonumber \\
g_{s}^j&=\beta_j  g_s,\qquad j=t,h,d,\alpha\\
g_{\rho}^i&=\delta_i g_\rho,\qquad i=t,\,h \nonumber,\\
g_{\rho}^i&=0,\qquad i=d,\,\alpha\nonumber,
\end{align}
where the  parameters $\eta_j$, $\beta_j$ and $\delta_i$ are constants.
Next we determine the dissolution 
density  ($\rho_{dis}$)  as a function of these parameters 
 and study the dependence of $\rho_{dis}$ on the global proton
 fraction ($Y_p$).
 Experimental and  theoretical information will used to fix
the parameters $\eta_j$, $\beta_j$ and $\delta_i$.
\begin{figure}
\begin{tabular}{c}
\includegraphics[width=1.0\linewidth,angle=0]{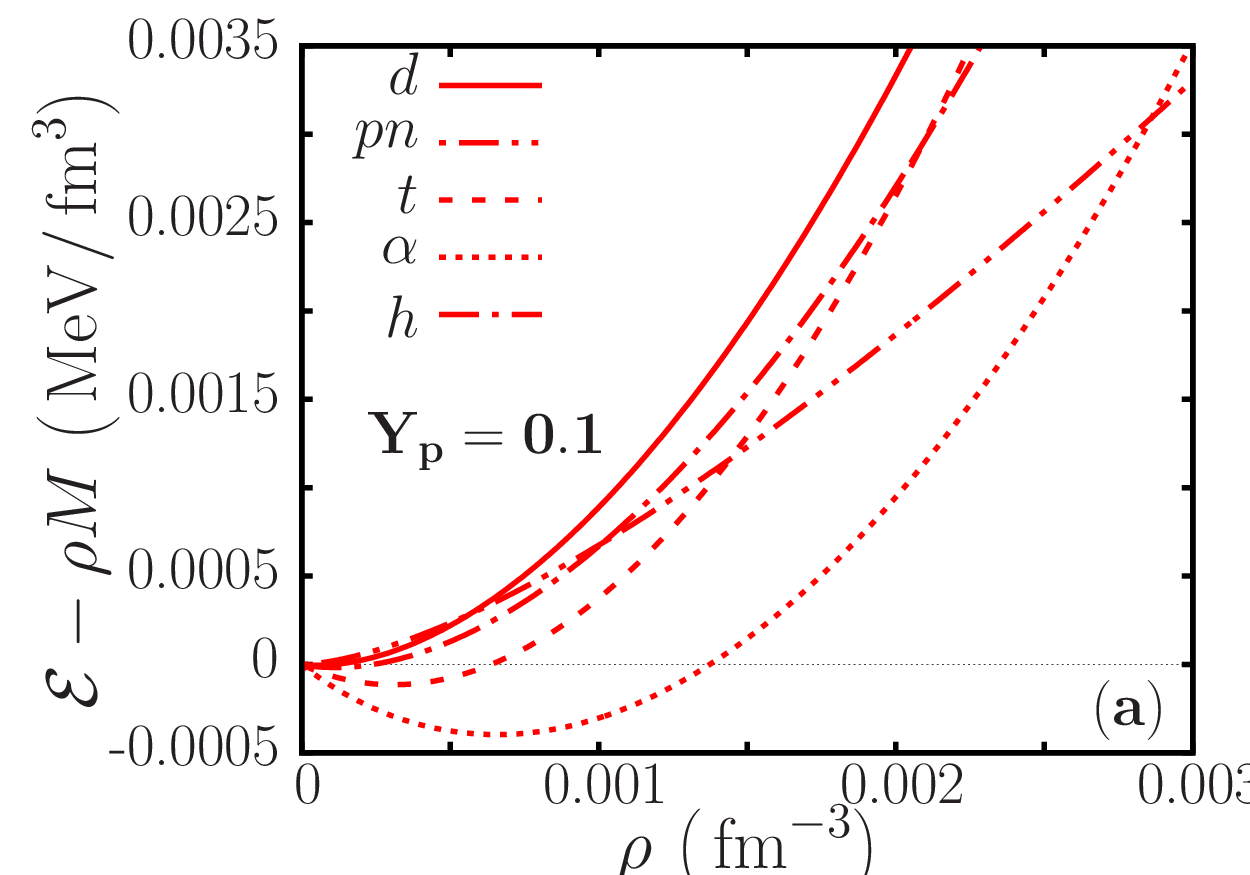}\\
\includegraphics[width=1.0\linewidth,angle=0]{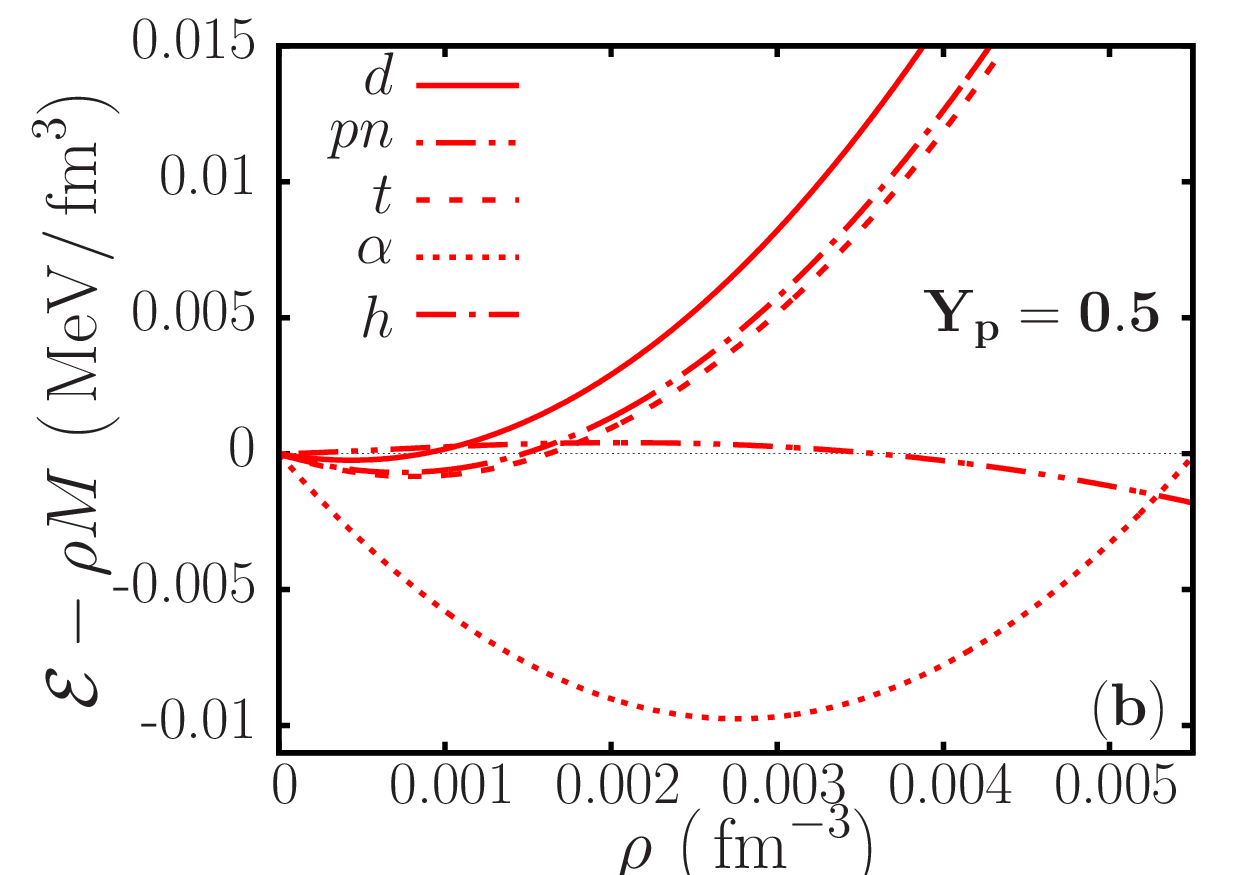}
\end{tabular}
\caption{(Color online) The free energy density minus the mass contribution  at $T=0$
  MeV as a function of the baryonic density for
$Y_p=0.1$ (a) and $Y_p=0.5$ (b) for the FSU parametrization, with  choice
(\ref{simple})  for the cluster couplings. The EOS
designed by the light cluster $i$ has the largest possible
fraction of  clusters $i$ and if, necessary, also protons and neutrons. }
\label{dens_energ}
\end{figure}

\subsection{Simple choice of the coupling parameters}

In order to have a feeling of the effect of including light clusters in the EOS,
we will first fix the cluster coupling constants as $g_s^i=0$, $g_v^i=A_ig_v$ for $i=t,h,d,\alpha$
 and $g_\rho^h=g_\rho^t=g_\rho$ by 
fixing 
\begin{eqnarray}
\eta_j&=&A_j,\,\,\, \beta_j=0,\, \quad j=d,\,t,\,h,\,\alpha, \nonumber\\
\delta_i&=&1 , \quad i=t,\,h, \label{simple}\\
 \delta_i&=&0 , \quad i=d,\,\alpha, \nonumber 
\end{eqnarray}
in (\ref{eq:cc}). This choice corresponds to the one used in \cite{alphas,clusters}
and may be justified in the following way: a) since we do not know
how nuclear matter
affects the binding energy of the light clusters we take the $\sigma$ coupling constants of the light 
clusters as zero ($g_s^i=0$); b) we expect that the $\omega$ meson will play the role of an excluded 
volume, i.e., at large densities will hinder the superposition of clusters by giving rise to a repulsive 
contribution. We take the coupling $g_v^i=A_ig_v$ and consider that
this would be of the order of magnitude expected; c) we consider that
the coupling of the light clusters to the $\rho$ meson
 is defined by the corresponding cluster isospin.


The dissolution density at a given proton fraction and zero temperature is determined 
from the crossing between the proton-neutron free 
energy and the free energy of matter with the largest number of clusters it is possible 
to form. This is the most stable matter with clusters.
 The stable phase is the one with the lowest free energy density. At low densities 
the equilibrium phase is the cluster phase and at large densities it is the proton-neutron 
matter. In fact, from the Fig. \ref{dens_energ} it is seen that  this transition is
  characterized by a negative curvature of the free energy density, and therefore,
it does not correspond to a thermodynamically stable phase transition. The
inclusion of larger clusters is necessary to realistically describe this
transition but this is not the aim of the present work.

\begin{figure*}
\begin{tabular}{ll}
\includegraphics[width=0.45\linewidth,angle=0]{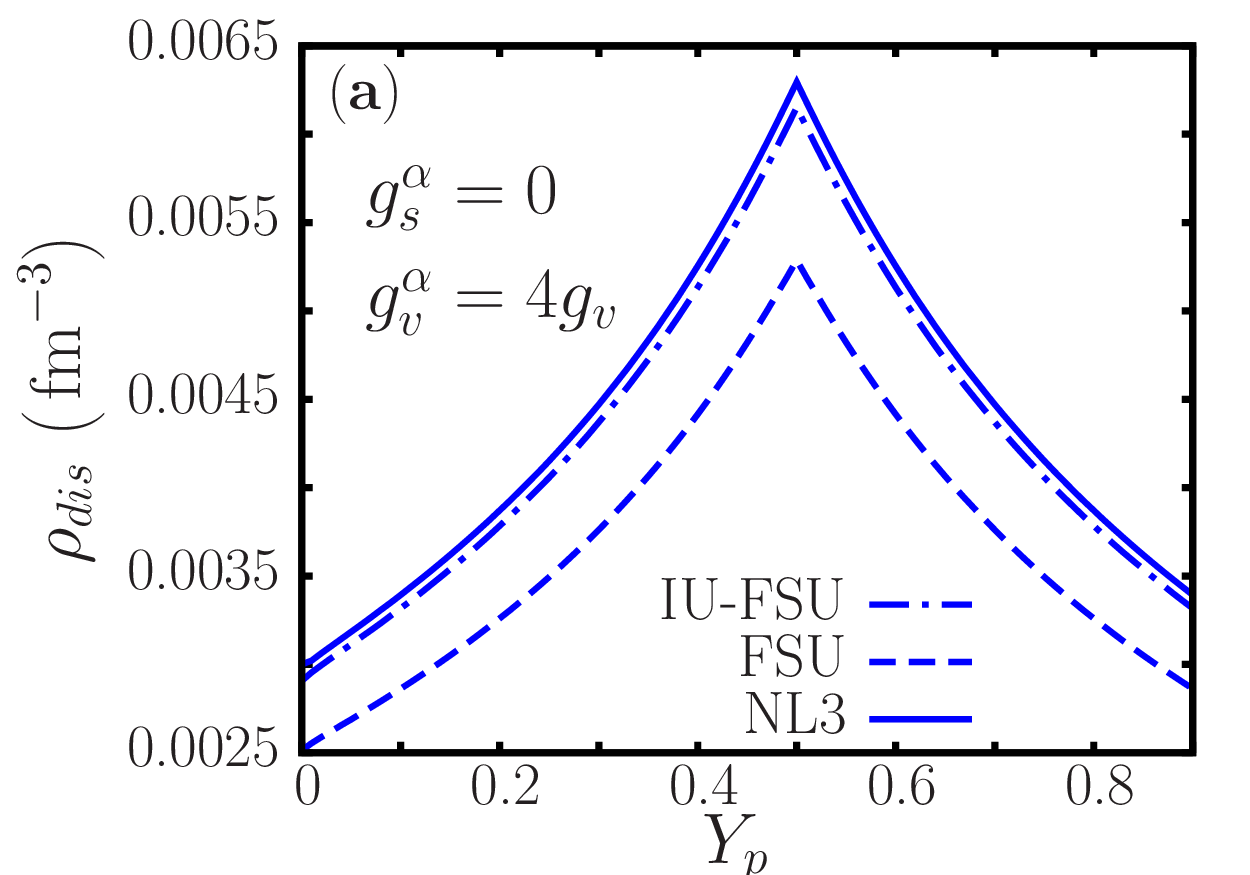}&
\includegraphics[width=0.45\linewidth,angle=0]{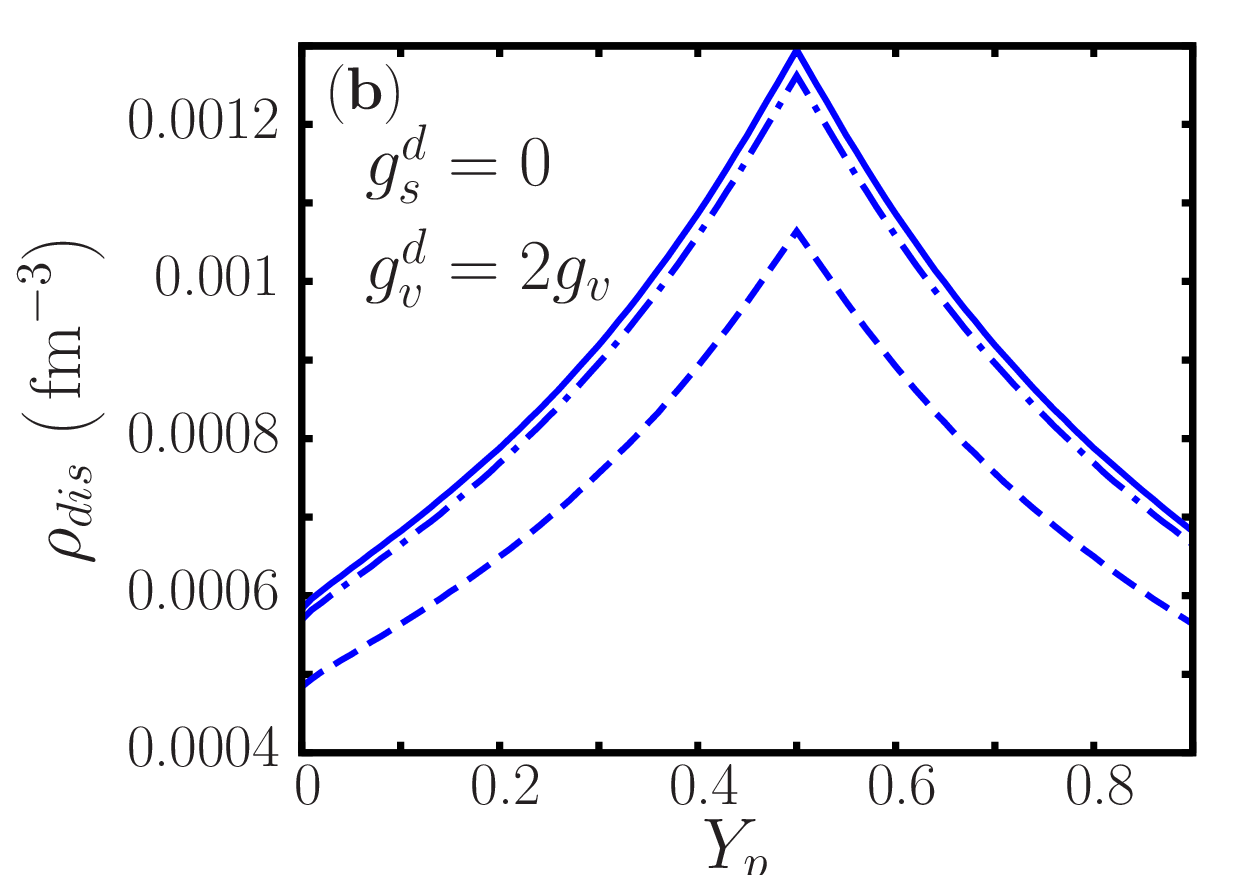}\\
\includegraphics[width=0.45\linewidth,angle=0]{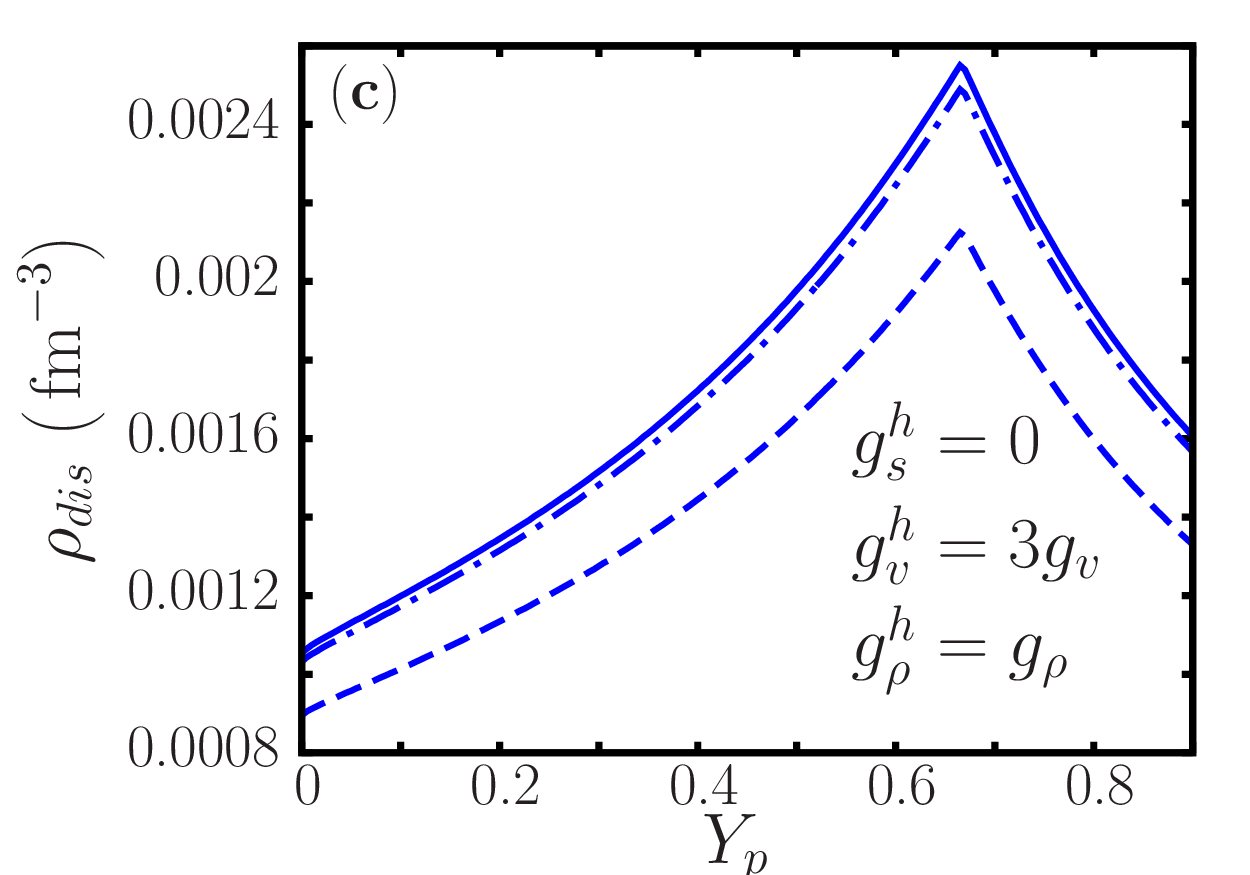}&
\includegraphics[width=0.45\linewidth,angle=0]{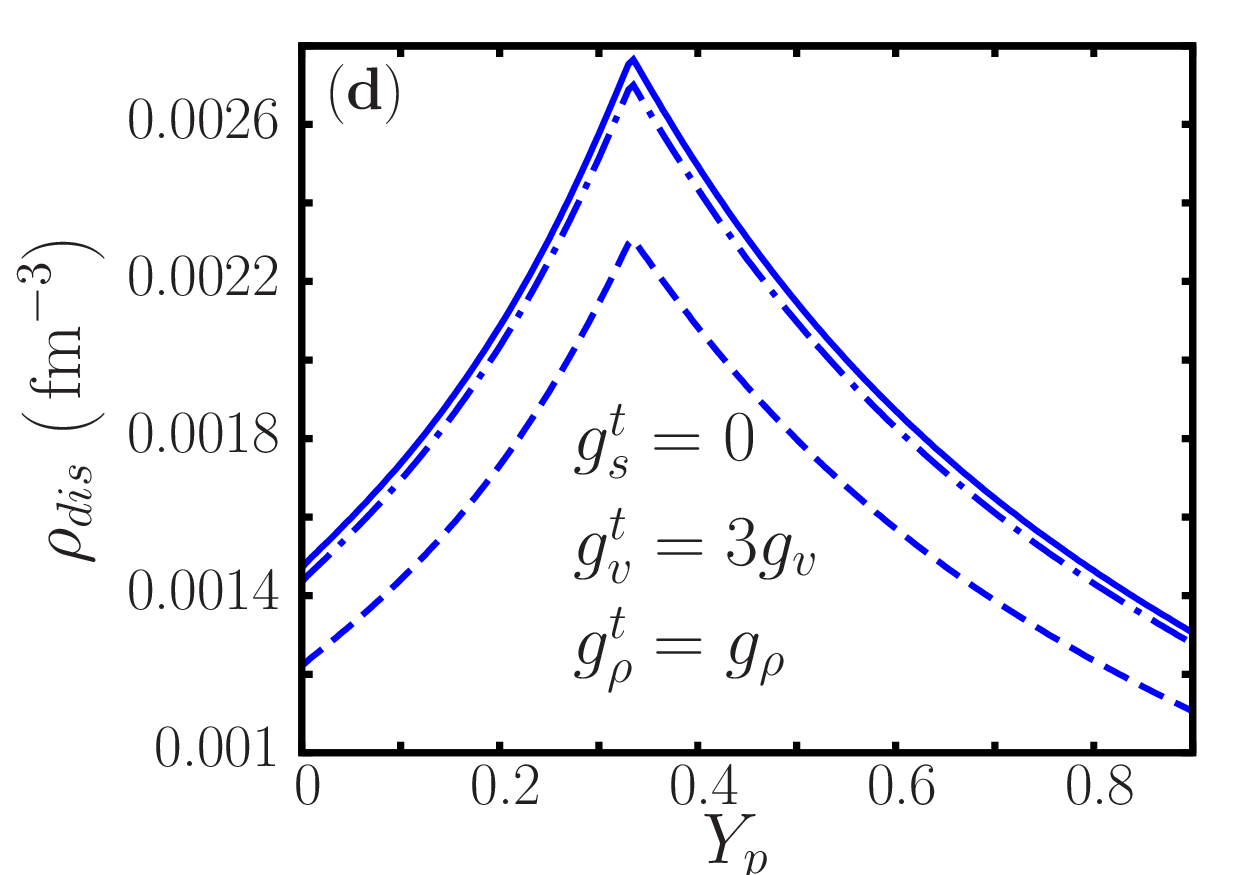}
\end{tabular}
\caption{(Color online) The dissolution density $\rho_{dis}$ for $T=0$ MeV as a function of total proton fraction $Y_p$ for the
various clusters, with  choice
(\ref{simple})  for the cluster couplings: $\alpha$ (a), deuterons (b), helions (c) and tritons (d). 
In each graph the curves represent NL3, IU-FSU and FSU from top to bottom, respectively.}
\label{fig:rho_d}
\end{figure*}
The free energy density minus the mass contribution, ${\cal F}=\mathcal{E}-M\rho$, 
is plotted in Fig. \ref{dens_energ} as a function of the baryon  density for different EOS,
namely the proton-neutron EOS without clusters and the proton-neutron matter with the largest possible fraction of one type of light
cluster for the proton fractions considered ($Y_p=0.1$ and $Y_p=0.5$). The
curves in Fig. \ref{dens_energ} were obtained for FSU but
for the other parametrizations the results are qualitatively the same. From Fig. \ref{dens_energ}
we see that there is some range of densities in which the free energy
density of the equations of state with one type of cluster is
lower than the pure nucleonic matter (without any bound states). In that range, the system can decrease its energy
density by forming clusters.
At $T=0$, the EOS with $\alpha$ particles is by far the most stable followed by the EOS with tritons, helions and deuterons. This
sequence of stability is explained by the binding energy per nucleon of each cluster, having the $\alpha$ particle the 
highest value and the deuteron the lowest one. 

 At finite temperature, all the clusters will be in chemical equilibrium and  their abundances will be determined by a balance between
temperature, their mass and binding energy.

The calculated dissolution densities are  shown in Fig. \ref{fig:rho_d}
for all the clusters and the three NLWM paramerizations with the
choice (\ref{simple}) for the cluster couplings. For the deuteron and $\alpha$ 
particles the maximum dissolution density is at $Y_p=0.5$ allowing
the conversion of all nucleons into bound states (deuteron or $\alpha$ particles).
At $Y_{p}=2/3$ and $Y_{p}=1/3$, all the nucleons are converted
into helions and tritons, respectively, giving the highest dissolution density.
The NL3 and IU-FSU parametrizations give similar dissolution
densities, with NL3 giving the highest ones. The IU-FSU
parametrization gives results between FSU and NL3, very close
to NL3, therefore, it will not be considered
in most of the discussion and only the NL3 and FSU will be shown in the figures.
The dissolution density of the light clusters within the different models is mainly determined by the
isoscalar properties of the model. In particular, FSU has a quite large coupling constant to the $\sigma$
meson which defines a smaller effective nucleon mass at low densities and, therefore, clusters will dissolve at smaller densities. 
This  is also the reason why the NL3 dissolution density is the largest: for the densities of interest NL3
has the smallest nucleon effective mass.

We next generalize the choice (\ref{simple}) varying smoothly one
parameter ($\eta_i,\, \beta_i,\, \delta_i$) each
 time and keeping the
other.
\begin{figure}
\begin{tabular}{c}
\includegraphics[width=1.0\linewidth,angle=0]{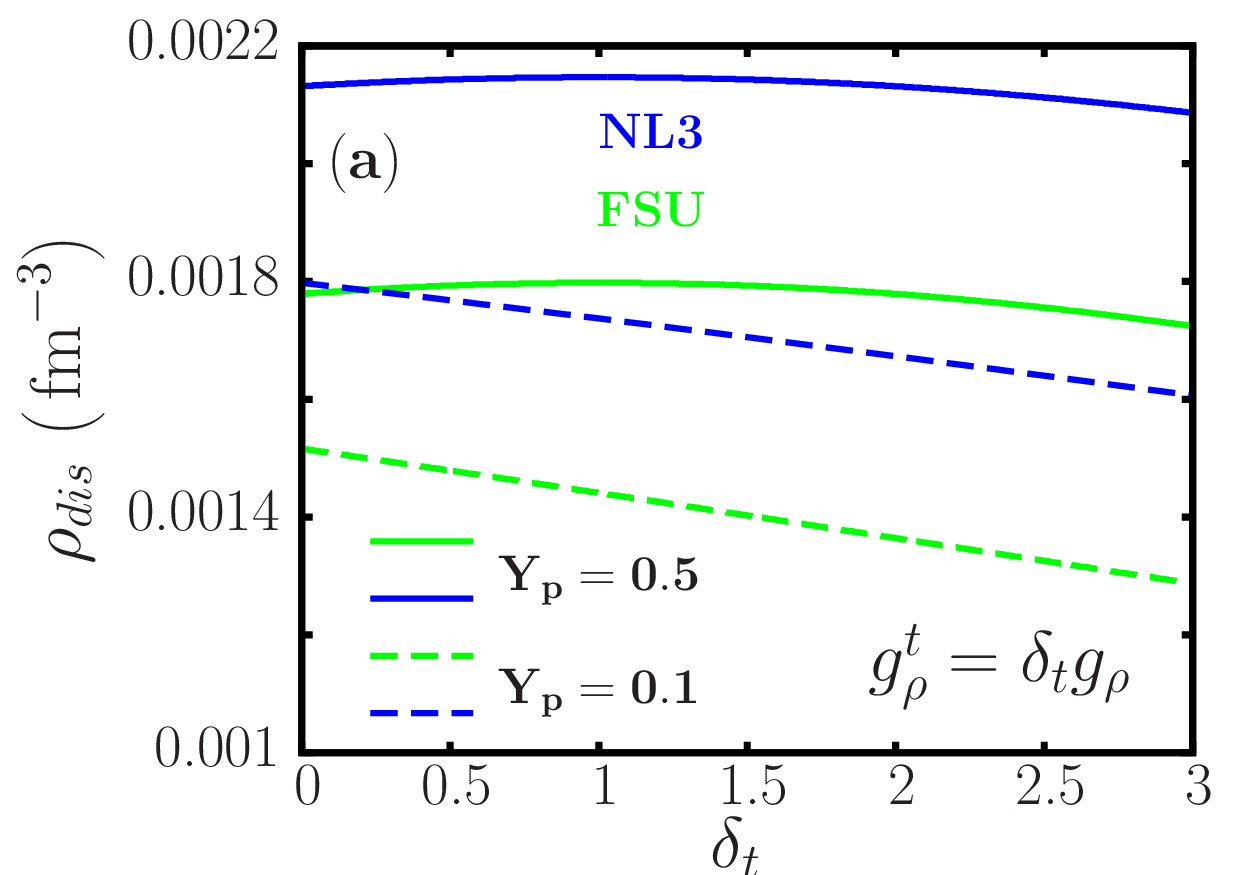} \\
\includegraphics[width=1.0\linewidth,angle=0]{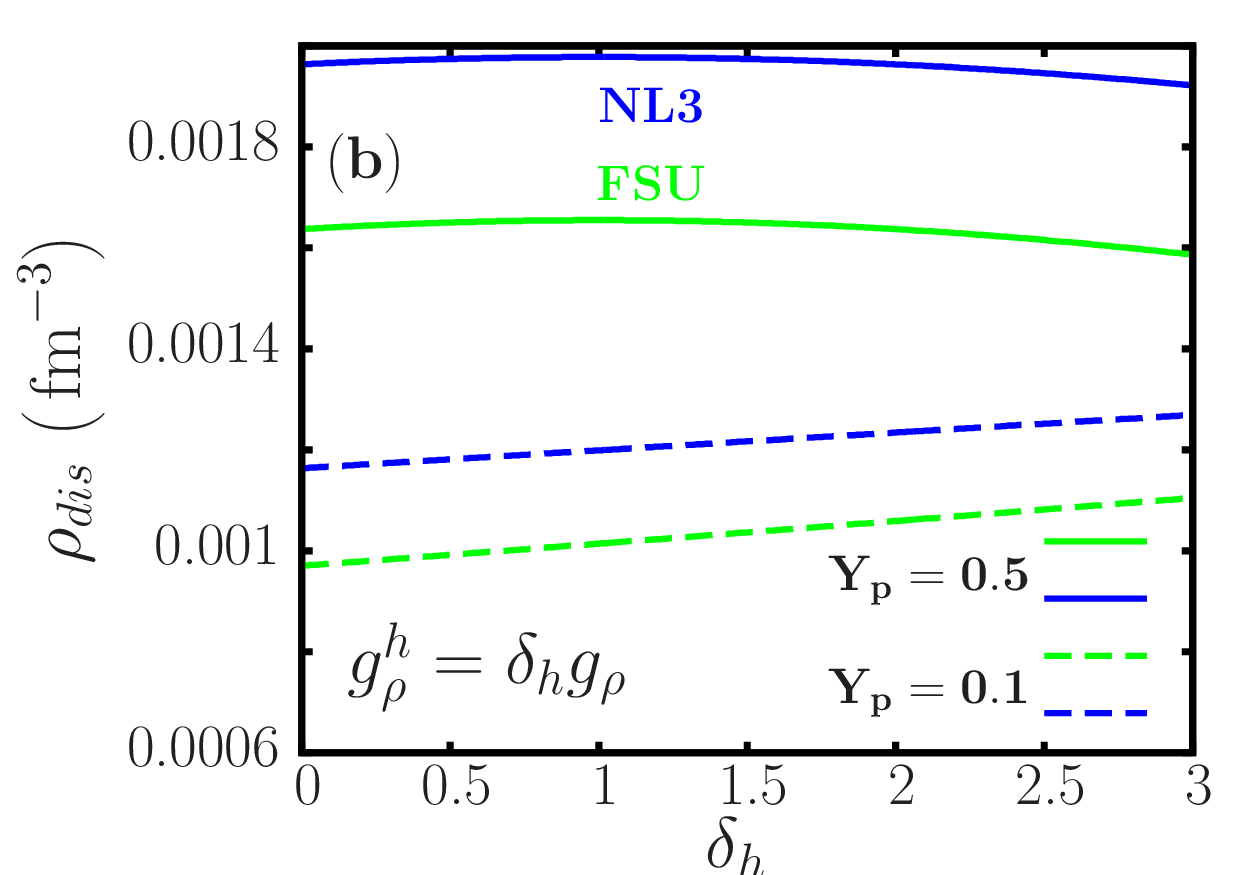}
\end{tabular}
\caption{(Color online) The dissolution density for tritons (a) and helions (b) as a function of $\delta^i$ ($g_\rho^i=\delta_i g_\rho$) for
$Y_p=0.5$ (solid lines) and $Y_p=0.1$ (dashed lines).}
\label{g_rho}
\end{figure}

\subsection{Varying the $g_\rho^i$ coupling constant}

To study the effect of the $\rho$ meson-cluster coupling constants $g_\rho^i$ on the dissolution density 
for the helion and triton, we fix the other coupling constants by $g_v^i=A_ig_v$ and $g_s^i=0$ and
parametrize  $g_\rho^i$ by $g_\rho^i=\delta_i g_\rho$ ($i=h,t$).
The results are shown in Fig. \ref{g_rho}.\\
The densities of all  species that are present are defined by the global proton
fraction $Y_p$. For values of $Y_p$ where only tritons and neutrons are present 
in nuclear matter, the free energy density term that depends on $g_\rho^t$ is
\begin{equation}
\label{eq:asym1}
 \mathcal{F}\pc{\delta_t}=\frac{1}{8}\frac{g_\rho^2}{\pc{m^*_\rho}^2}\pc{\delta_t\rho_t +\rho_n }^2
\end{equation}
and in the case where only helions and neutrons are present, 
\begin{equation}
\label{eq:asym2}
\mathcal{F}\pc{\delta_h}=\frac{1}{8}\frac{g_\rho^2}{\pc{m^*_\rho}^2}\pc{\delta_h\rho_h-\rho_n }^2.
\end{equation}
where $\pc{m_\rho^*}^2=m_\rho^2+2g_\rho^2g_v^2\Lambda_v\pc{\omega^0}^2$.\\
Thus, for $Y_p=0.1$, an increase in $g_\rho^h$ through the parameter $\delta_h$,
reduces the symmetry energy
of a system composed by helions and consequently, the dissolution density increases. For a system
made of tritons the opposite happens, and an increase in $g_\rho^t$ increases
the system symmetry energy leading to a decrease in the dissolution density.\\
For $Y_p=0.5$, the helion and triton systems have a similar behavior, at some values of $g_\rho^t$ 
and $g_\rho^h$ they acquire the lowest free energy density, and consequently, the highest dissolution 
density due to the vanishing of the symmetry term, (\ref{eq:asym1}) or (\ref{eq:asym2}).

One important feature  seen in  Fig. \ref{g_rho} is the small range of variation of the dissolution 
density as a function of $g_\rho^h$ and $g_\rho^t$, compared with the range of variation of 
the dissolution density as a function of the $\omega$ and $\sigma$ cluster coupling
constants that we will discuss next. The cluster formation and dissolution is quite insensitive to the
$\rho$ meson-cluster coupling constant.

\subsection{Varying the $g_v^i$ coupling constant}
\begin{figure*}
\begin{tabular}{ll}
\includegraphics[width=0.45\linewidth,angle=0]{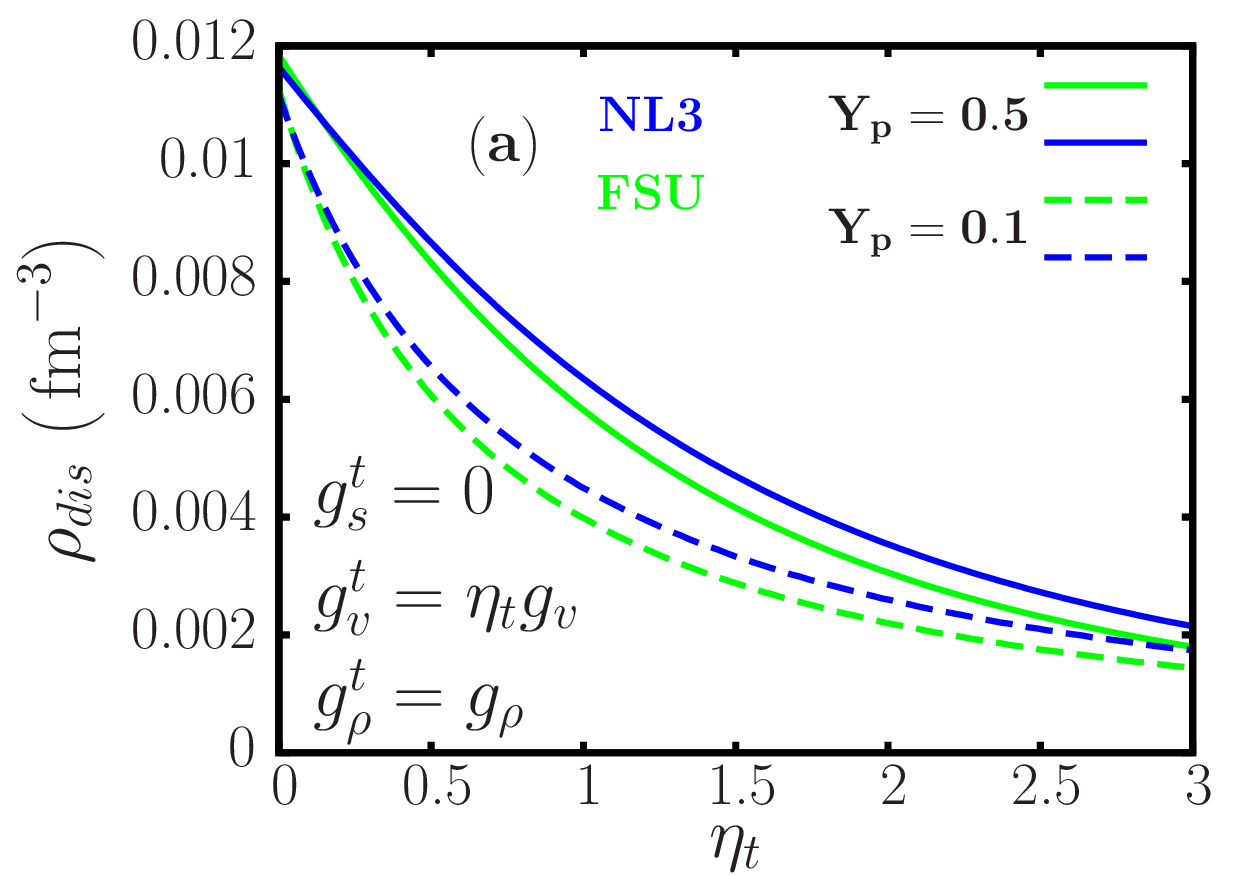}&
\includegraphics[width=0.45\linewidth,angle=0]{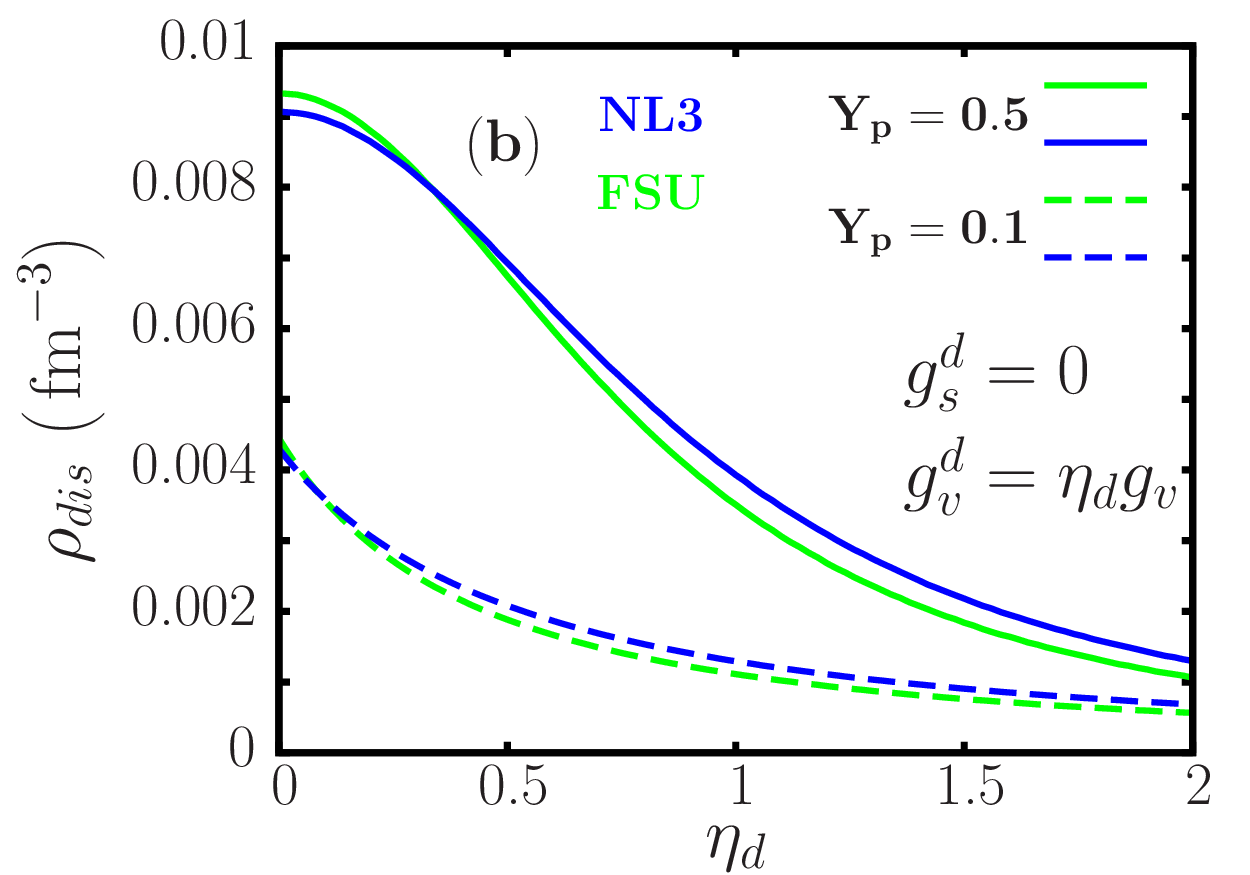}\\
\includegraphics[width=0.45\linewidth,angle=0]{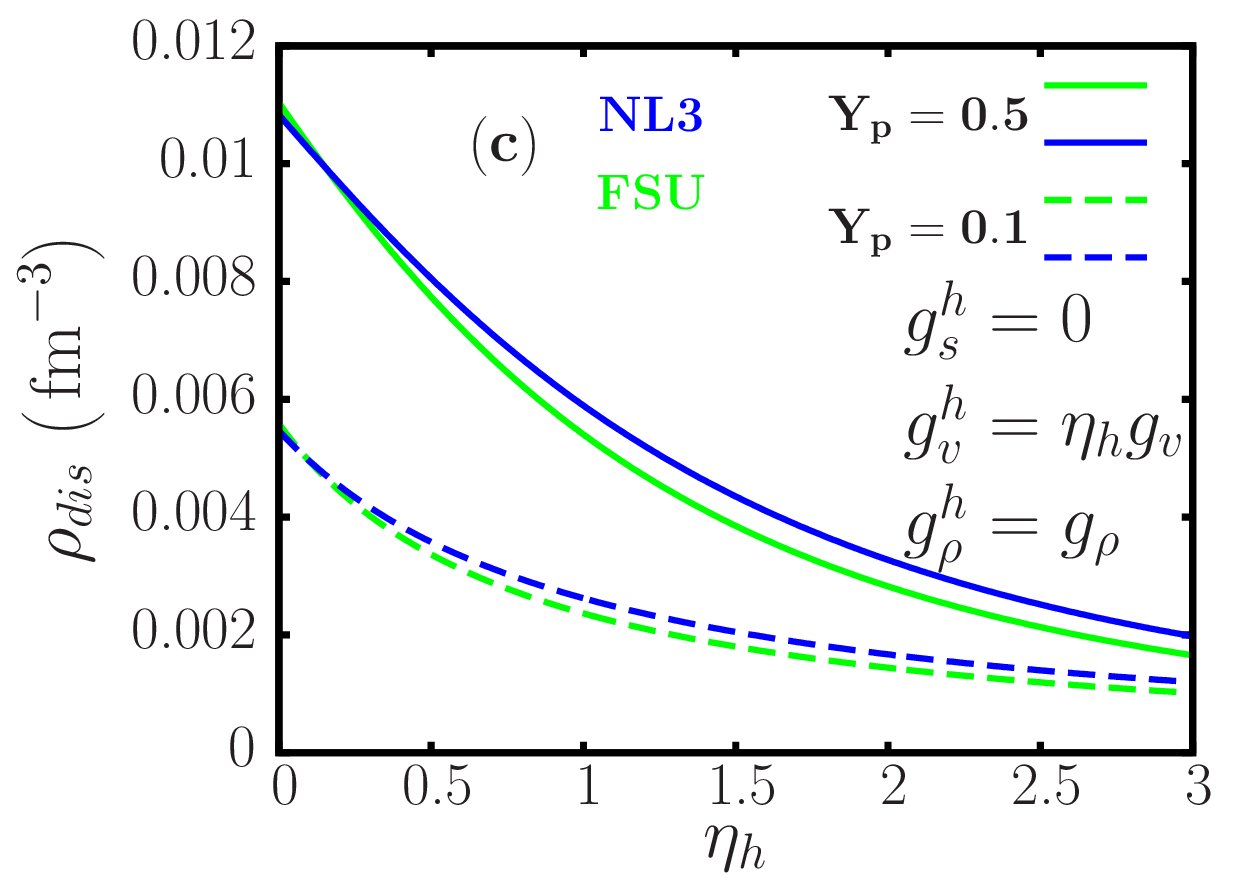}&
\includegraphics[width=0.45\linewidth,angle=0]{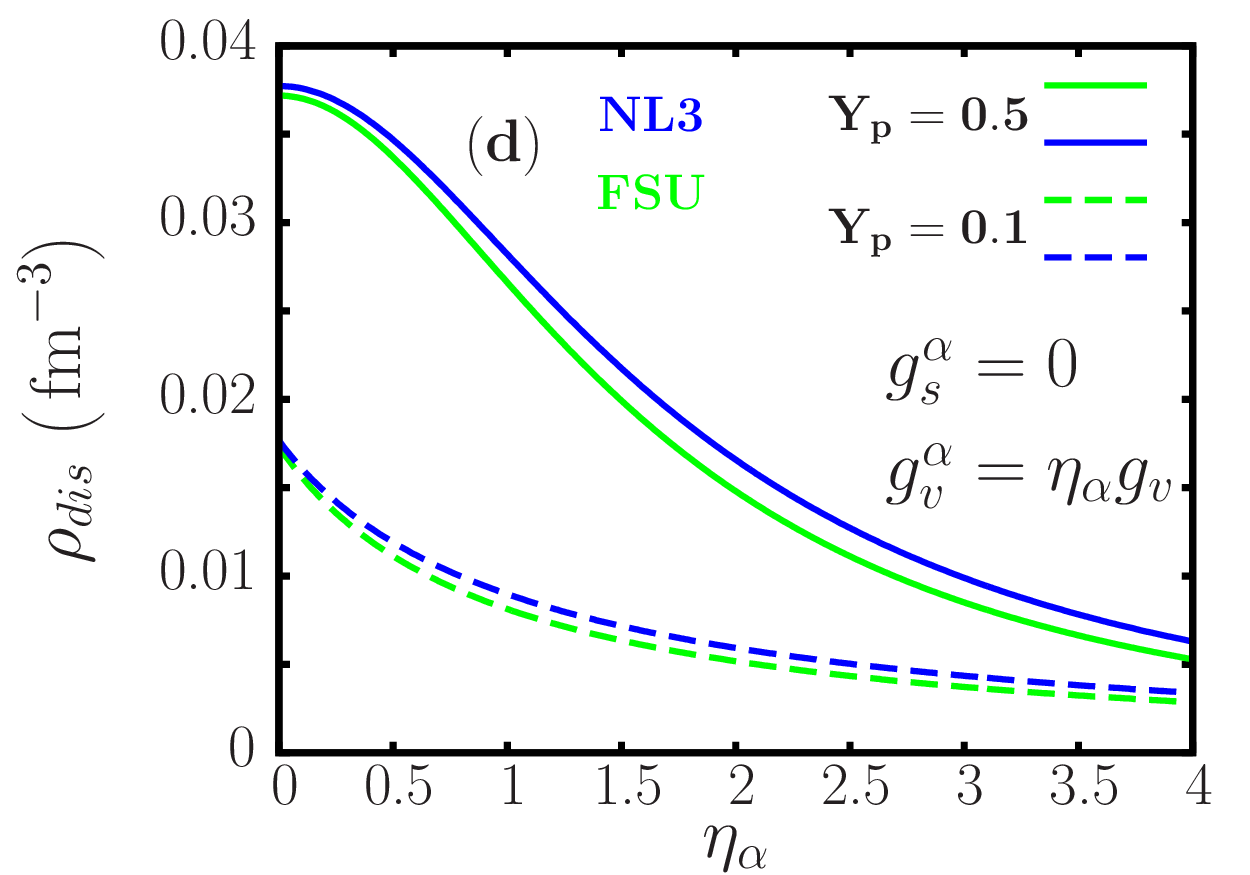}
\end{tabular}
\caption{(Color online) Dissolution density of each cluster as a function of $\eta_i$ ($g_v^i=\eta_i g_v$) for
$Y_p=0.5$ (solid lines) and $Y_p=0.1$ (dashed lines): tritons (a), deuterons (b),
helions (c) and $\alpha$ (d).}
\label{rho_v}
\end{figure*}
The effect of the $\omega$-cluster coupling constants on the dissolution
density is studied by
fixing the $\sigma$ and $\rho$ coupling constants as $g_s^i=0$ ($i=t,h,d,\alpha$) and
$g_\rho^i=g_\rho$ ($i=t,h$). The $g_v^i$ is parametrized by $\eta_i$ as $g_v^i=\eta_i g_v$. 
The dissolution density versus $\eta_i$ for all the  clusters is shown in Fig. \ref{rho_v}
for symmetric ($Y_p=0.5$) and asymmetric ($Y_p=0.1$) nuclear matter.

An increase in $g_v^i$  decreases the dissolution density for all clusters.
This is due to the fact that an increase in $g_v^i$ leads to a more
intense repulsion between the clusters, thus favoring their dissolution.

For $Y_p=0.1$ the dissolution density is lower than for $Y_p=0.5$. This was expected because for $Y_p=0.1$
the number of clusters that can be formed is smaller than for symmetric nuclear matter. To dissolve
a bigger number of clusters we need a bigger amount of energy and,
thus, makes the cluster states with $Y_p=0.5$ more
stable than $Y_p=0.1$. 

\subsection{Varying the $g_s^i$ coupling constant}
\begin{figure*}
\begin{tabular}{ll}
\includegraphics[width=0.45\linewidth,angle=0]{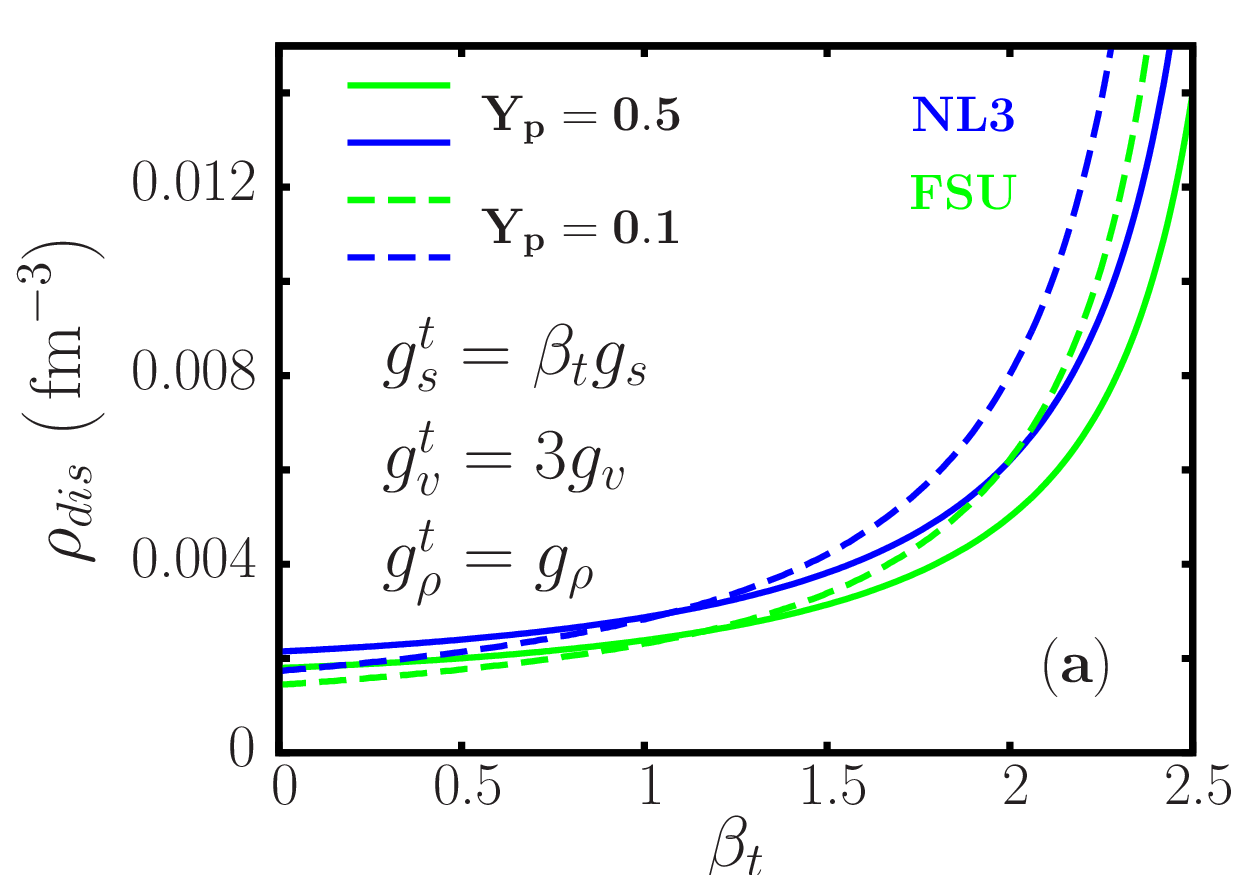}&
\includegraphics[width=0.45\linewidth,angle=0]{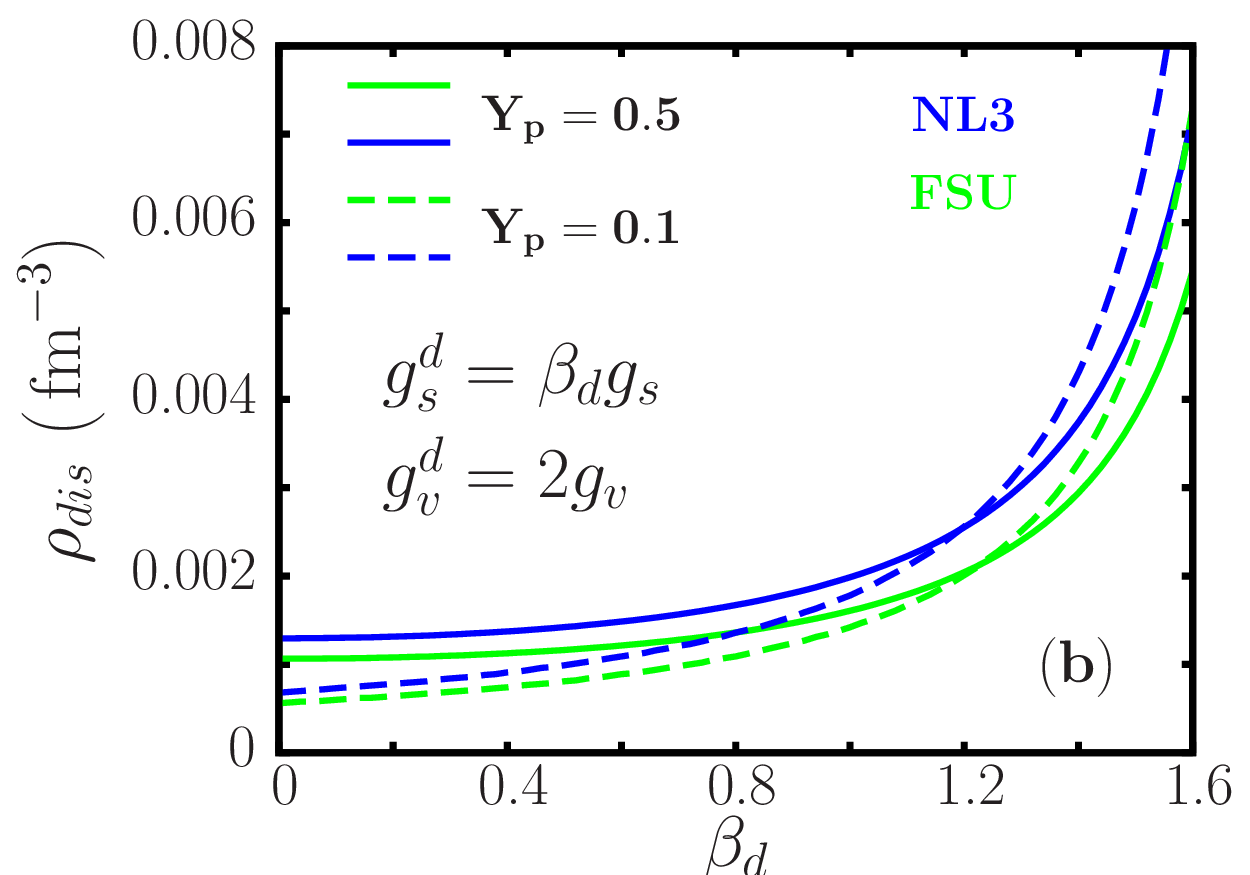}\\
\includegraphics[width=0.45\linewidth,angle=0]{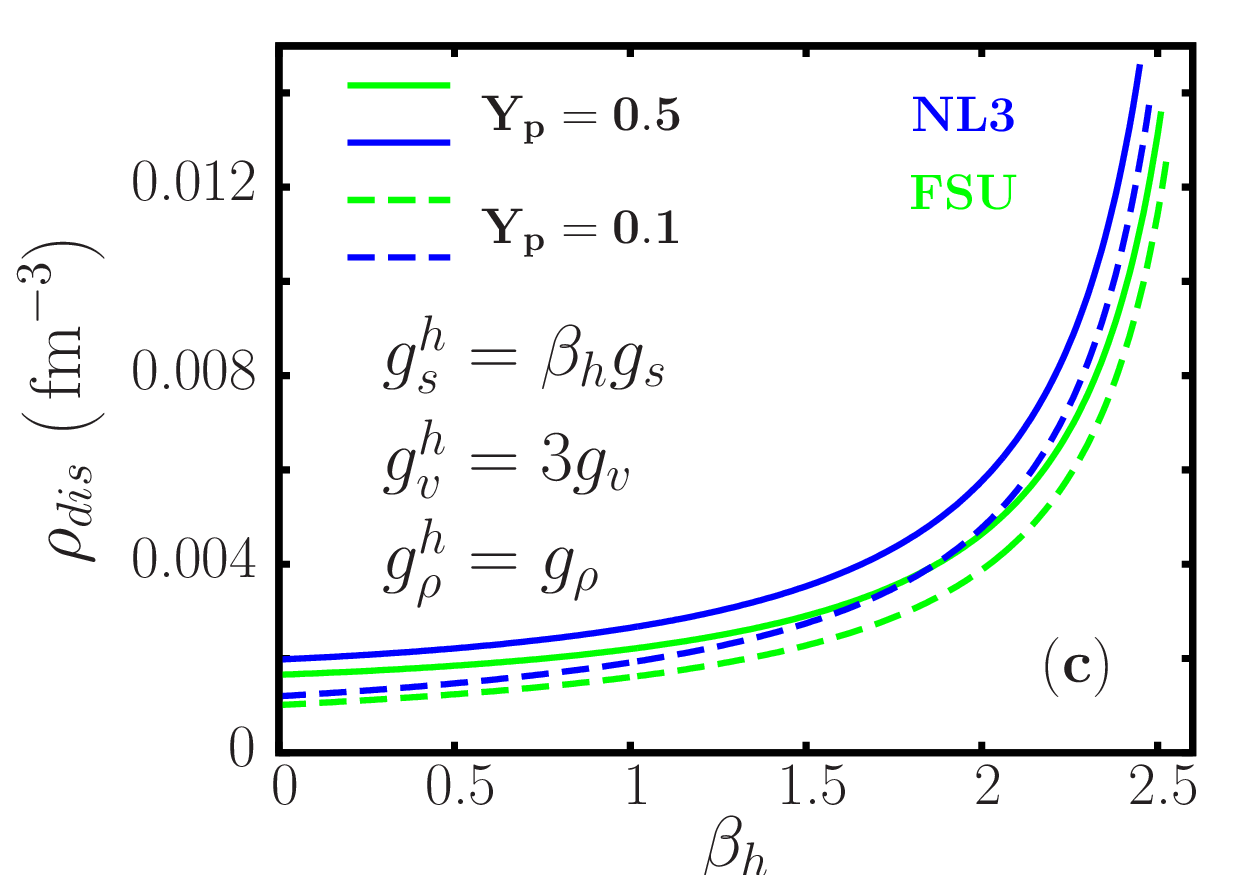}&
\includegraphics[width=0.45\linewidth,angle=0]{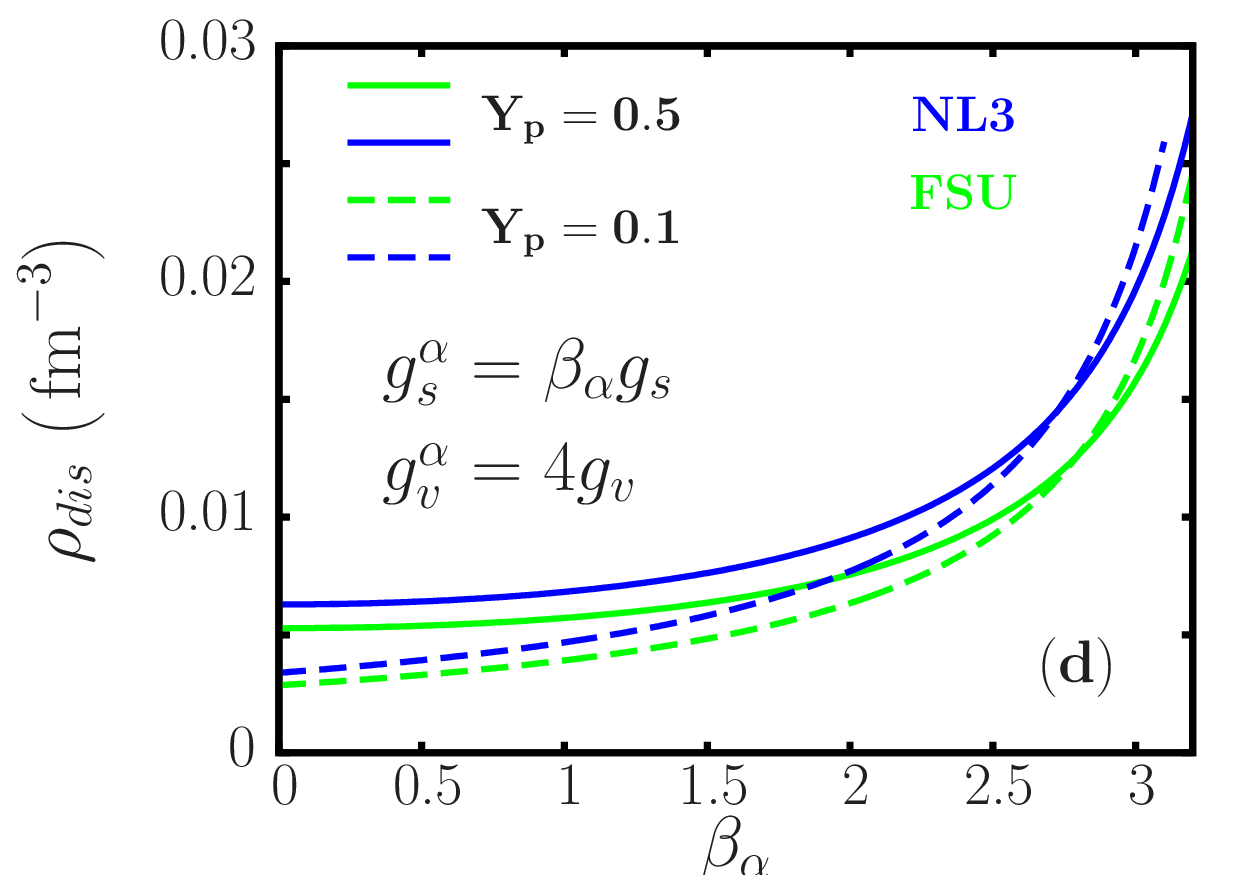}
\end{tabular}
\caption{(Color online) Dissolution density of each cluster as a function of $\beta_i$ ($g_s^i=\beta_i g_s$) for
$Y_p=0.5$ (solid lines) and $Y_p=0.1$ (dashed lines): tritons (a), deuterons (b),
helions (c) and $\alpha$ (d).}
\label{rho_s}
\end{figure*}
To study the dissolution density dependence on the $\sigma$-cluster coupling constants of all clusters $g_s^i$
we fix the $\omega$ and $\rho$ coupling constants as $g_\rho^i=g_\rho$ ($i=t,h$) and $g_v^i=A_ig_v$
($i=t,h,d,\alpha$) and parametrize $g_s^i$ as
$g_s^i=\beta_i g_s$.

The results are shown in Fig. \ref{rho_s} for $Y_p=0.5$ and $Y_p=0.1$. As expected the dissolution density grows 
as the $g_s^i$ increases. An increase in $g_s^i$ makes the cluster more bound (smaller effective mass) and,
therefore, cluster states become more stable. Except for the helion, in all other cases for a given value of $\beta$,
the dissolution densities of the clusters become larger for $Y_p=0.1$ than $Y_p=0.5$. This is because when considering
$Y_p=0.5$ we are comparing matter with clusters with nuclear symmetric matter, that is nuclear matter in its most
bound state. For $Y_p=0.1$ nuclear matter is not even bounded. Increasing the binding energy of the clusters will stabilize
them more relatively to the homogeneous asymmetric nuclear matter than in the previous case to symmetric matter. For helions
there are too few clusters in matter with $Y_p=0.1$ and therefore the curves $Y_p=0.5$ and $0.1$ will never cross. 

When $\beta_i$ is well below $A_i$, the dissolution density slightly changes for all clusters. For 
some value of $\beta_i$ below $A_i$ the dissolution density increases abruptly and the system becomes absolutely stable,
in other words, the clusters will not dissolve for any density.

\begin{figure}
\begin{tabular}{ll}
\includegraphics[width=0.9\linewidth,angle=0]{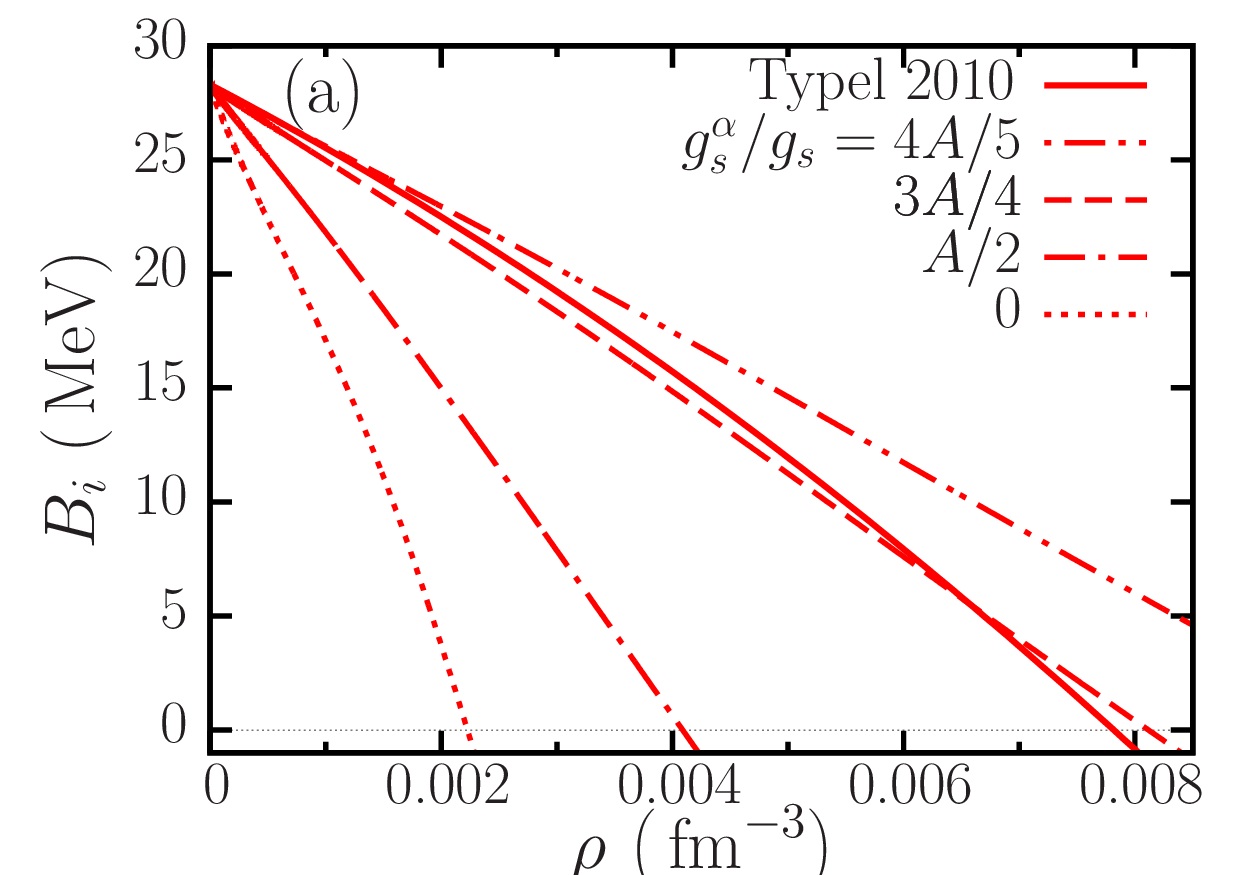}\\
\includegraphics[width=0.9\linewidth,angle=0]{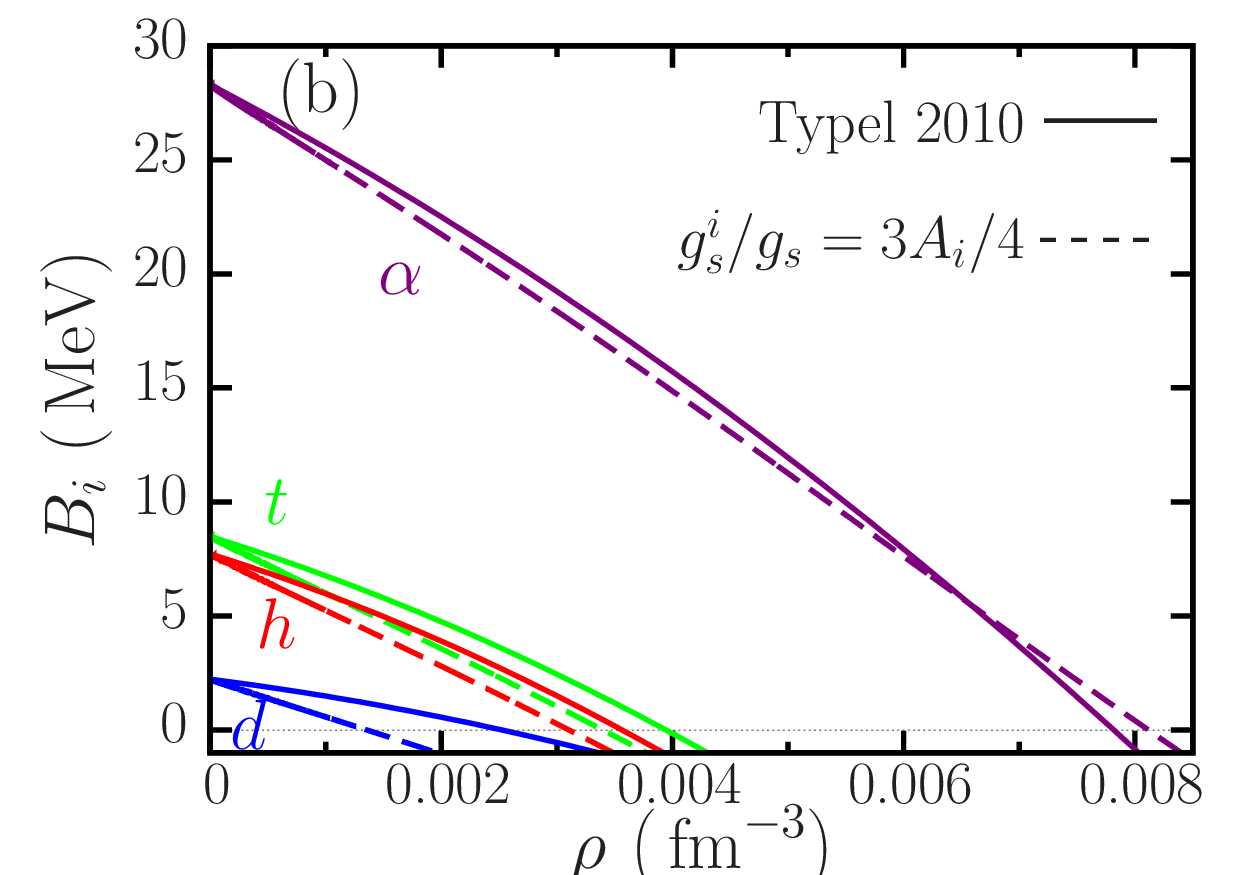}
\end{tabular}
\caption{(Color online) Binding energy, as defined in Eq. (\ref{bind}), of light clusters in
  equilibrium with symmetric  nuclear matter described within FSU at  $T=5$
  MeV.  In (a) the binding energy of $\alpha$-particles is given for several
  values of the coupling $g_{s\alpha}$ (dashed, dotted and dash-dotted
  curves). The $\alpha$ in-medium binding energy given in Typel 2010 \cite{typel2010}  
for $T=5$ MeV is also included (full line).  In (b) we consider
$g_{s}^i=3\, A_i/4\, g_s$ for all the clusters (dashed lines) and compare with the corresponding
results of Typel 2010 \cite{typel2010} at $T=5$ MeV (full lines).}
\label{mott}
\end{figure}

\subsection{A physical choice for the cluster coupling constants}\label{results4}

After analyzing the effect of the different cluster-meson couplings,
we discuss how these couplings could be fixed taking into account
theoretical many-body calculations and experimental information.

To fix the $\sigma$ and $\omega$ clusters couplings
 two constraints are required  for each cluster.
We  consider  the 
dissolution densities for symmetric nuclear matter { at $T=0$ MeV}
given in Ref. \cite{ropke2009}
and shown in Table \ref{tab:resul_energ}. 
However, this information is not enough to completely fix the cluster-meson couplings
and  we can find several sets of coupling constants that reproduce the dissolution 
densities. Therefore, we choose to  fix two coupling constants for each cluster and 
fit the remaining to reproduce its dissolution density, 
namely, we fix $g_s^i$ and $g_\rho^i$. For $g_s^i$ we take several values,
$0, \,A_i/2 \, g_s,\,3A_i/4 \, g_s, \,4A_i/5 \, g_s $, for all clusters
and for  $g_\rho^i$ we consider $g_\rho^i=g_\rho$, $\delta_i=1$, for triton and helion, and
$g_\rho^i=0$, $\delta_i=0$, for the isospin symmetric clusters, which is a
reasonable choice, and, as we have shown, this coupling does not affect much
the dissolution density of the cluster. Then, for each choice of
$g_s^i$,  $g_v^i$ is fit to reproduce the dissolution density of each cluster. The
 calculated values are shown in Table \ref{tab:fit}.
Different choices of $g_s^i$,  together with the above restriction for  $g_v^i$,
will allow us to discuss the effect of this
coupling constant on the overall particle fraction. In general we should have
a different ratio $g_s^i/g_s$ for each cluster, but, to simplify the
discussion, we take the same for all clusters.

\begin{figure}
\begin{tabular}{ll}
\includegraphics[width=0.9\linewidth,angle=0]{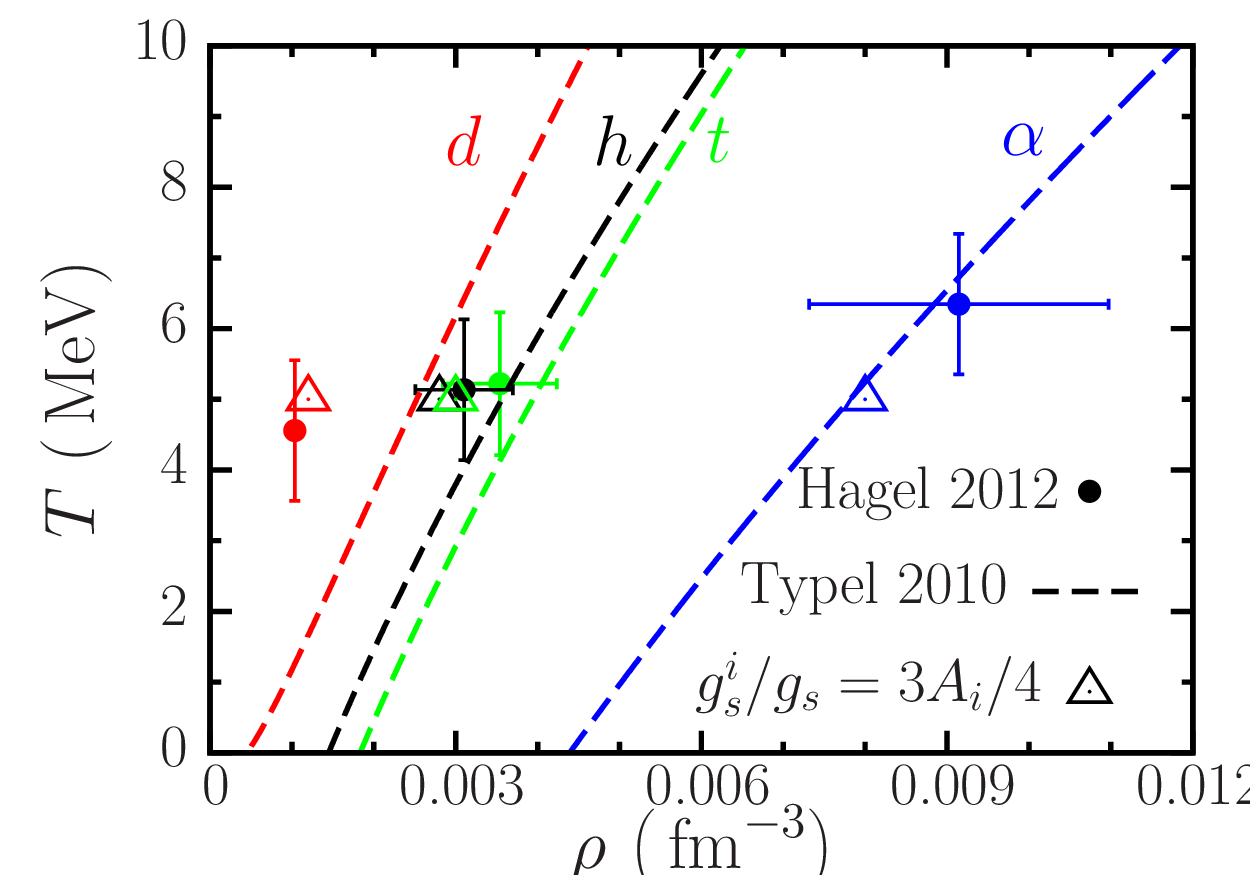}\\
\end{tabular}
\caption{(Color online) The Mott density  for different
  temperatures and light clusters: results from   Typel 2010 \cite{typel2010} 
(dashed lines), experimental prediction by Hagel 2012 \cite{hagel2012} (full dots with
errorbars), results for $g_s^i/g_s=3\, A_i/4$ and $T=5$
MeV (open triangle).}
\label{exp}
\end{figure}

\begin{table}
\renewcommand{\arraystretch}{1.2}
\setlength{\tabcolsep}{10pt}
\caption{Cluster coupling constants $g_v^i$ obtained for symmetric matter ($Y_p=0.5$)
that reproduce the dissolution densities of Table \ref{tab:resul_energ}, for
several values of the coupling $g_s^i$. $g_\rho^i/g_\rho=1$ for the triton and the
helion and zero for the  deuteron and the $\alpha$. }
\begin{tabular}{cccccc}
\hline
\hline
\multicolumn{6}{c}{$g_v^i/g_v$}\\
\hline
$g_s^i/g_s$& $0$ & $A_i/2$& $2A_i/3$ & $3A_i/4$&  $4A_i/5$\\
\hline
FSU      &           	&	 &		&	&		\\
$t$       & $2.961$  &  $3.793$ &  $4.177$ &  $4.382$ & $4.509$ \\
$h$       & $3.283$  &  $4.064$ &  $4.428$ &  $4.624$ & $4.745$ \\
$d$       & $3.038$  &  $3.260$ &  $3.422$ &  $3.516$ & $3.577$ \\
$\alpha$  & $4.440$  &  $5.027$ &  $5.439$ &  $5.675$ & $5.824$ \\
\hline
NL3      &            	&	 &		&	&		\\
$t$  & $3.357 $&   $4.172$&   $4.559$&   $4.765$&   $4.892$\\
$h$  & $3.718 $&     $4.485$&   $4.847$&   $5.043$&   $5.164$\\
$d$  & $3.371$&    $3.585$&   $3.744$&   $3.837$&   $3.897$\\
$\alpha$  & $4.919$&    $5.491$&   $5.897$&   $6.131$&   $6.278$\\
\hline
IUFSU     &           	&	 &		&	&		 \\
 $t$  & $3.303 $ &   $4.126$ &  $4.508$ &   $4.715$&   $4.842$\\
$h$  & $3.658$ &   $4.429$ &  $4.792$ &  $4.988$&   $5.110$\\
$d$  & $3.327$ &  $3.543$ &  $3.702$ &  $3.796$&   $3.859$ \\
$\alpha$  & $4.854$  & $5.429$ &  $5.837$ &  $6.071$&   $6.220$\\
\hline
\hline
\end{tabular}
\label{tab:fit}
\end{table}

The relative magnitude of the $\omega-$cluster coupling of the different models
is related to the $\omega-$nucleon coupling constant $g_v$: the smaller the values
of $g_v$ the larger the $\omega-$cluster coupling fractions
$g_v^i/g_v$, and 
the difference between $g_v^i$ for the models used is around $1\%$. In
fact, for example, we find that for $g_s^i=0$,  $g_v^t$ is
$43.22$ (NL3), $42.40$ (FSU) and $42.92$ (IU-FSU).

More experimental information is needed in order to determine the coupling
constants of the phenomenological model we propose. Recently, 
in-medium binding energies and Mott points of $d$, $t$, $h$ and $\alpha$
clusters have been determined experimentally \cite{hagel2012}. For
each cluster a combination of density and temperature corresponding to a
vanishing  in-medium cluster binding
energy, the Mott point,
were determined in a temperature range close to $T=5$ MeV. These
results, together with the dissolution densities at $T=0$ MeV given in
Table \ref{tab:resul_energ},  will
be used to determine the cluster coupling constant $g_s^i$.

\begin{figure*}
\begin{tabular}{ll}
\includegraphics[width=0.45\linewidth,angle=0]{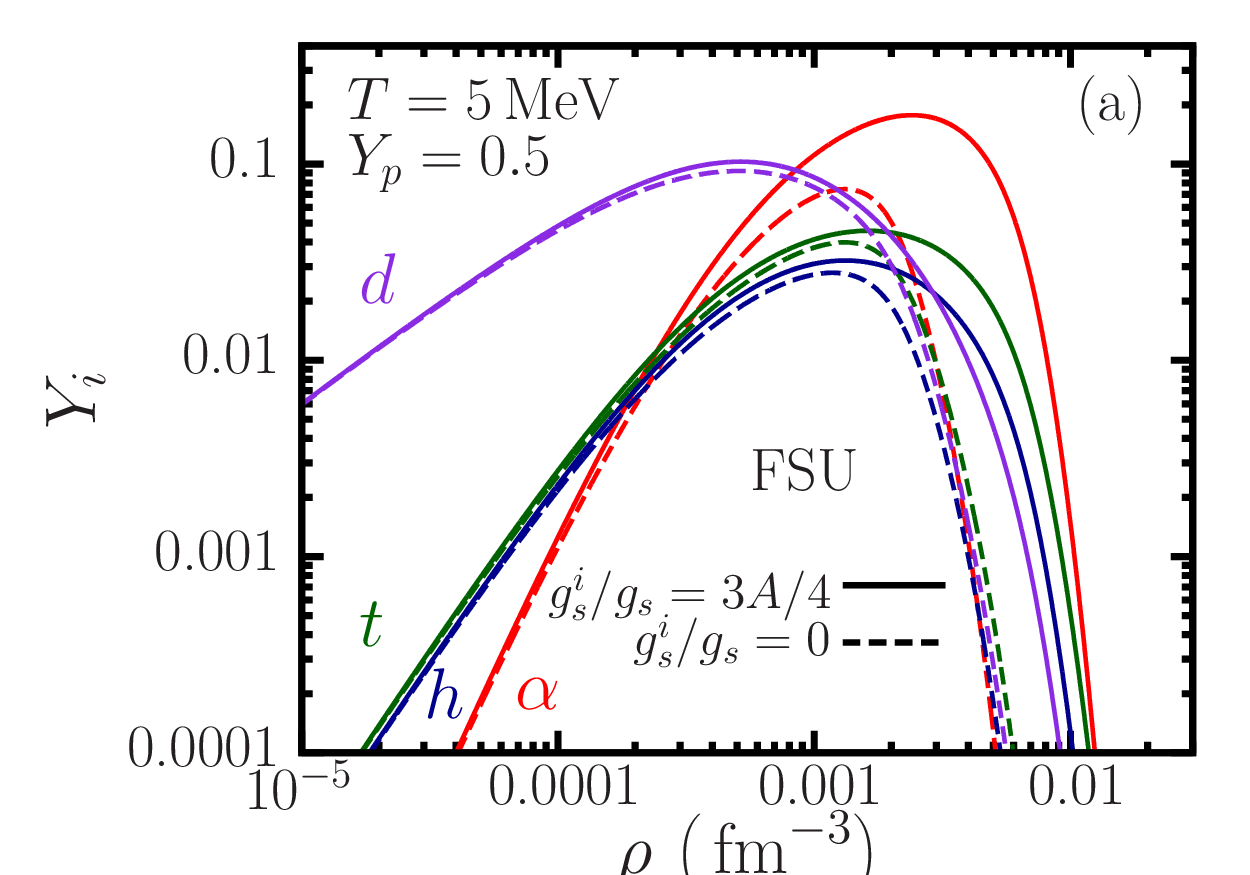}&
\includegraphics[width=0.45\linewidth,angle=0]{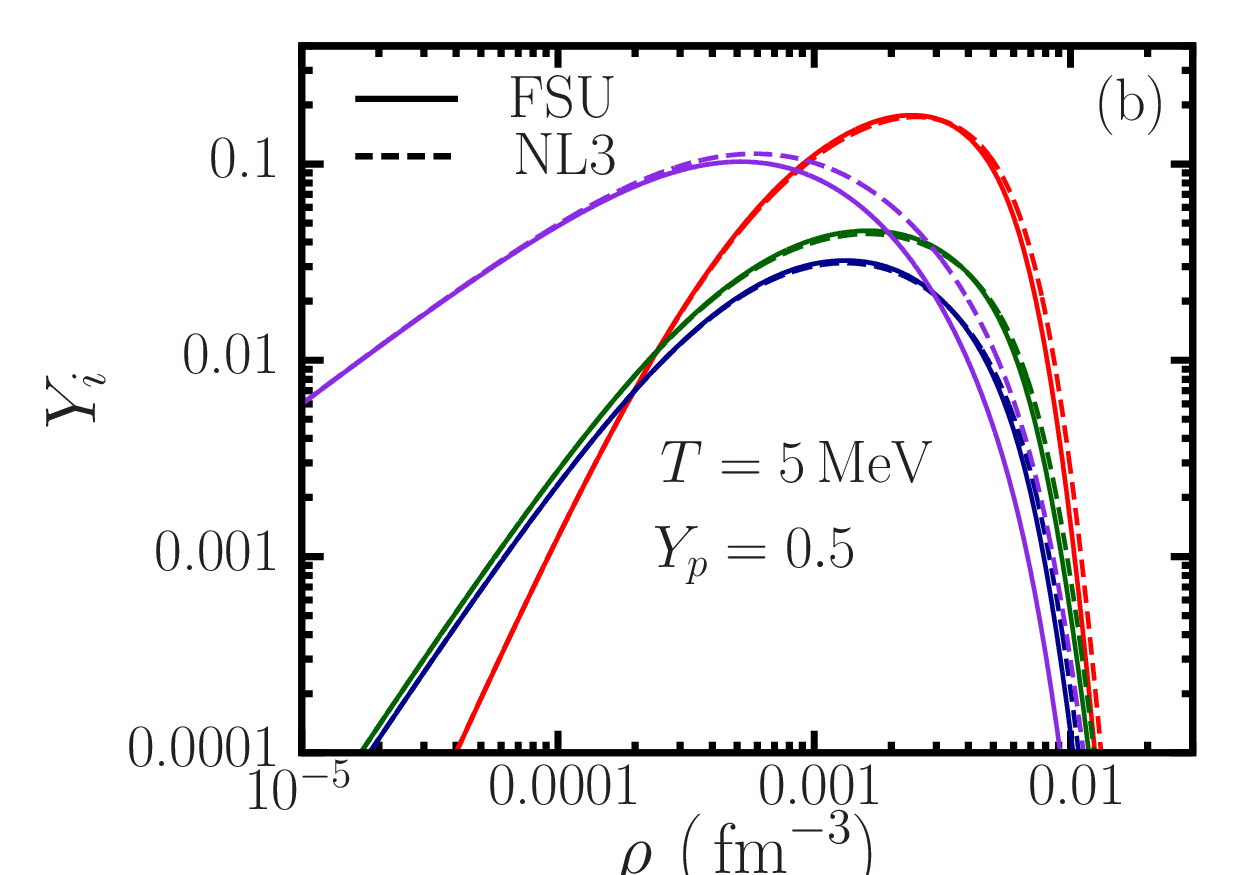}\\
\includegraphics[width=0.45\linewidth,angle=0]{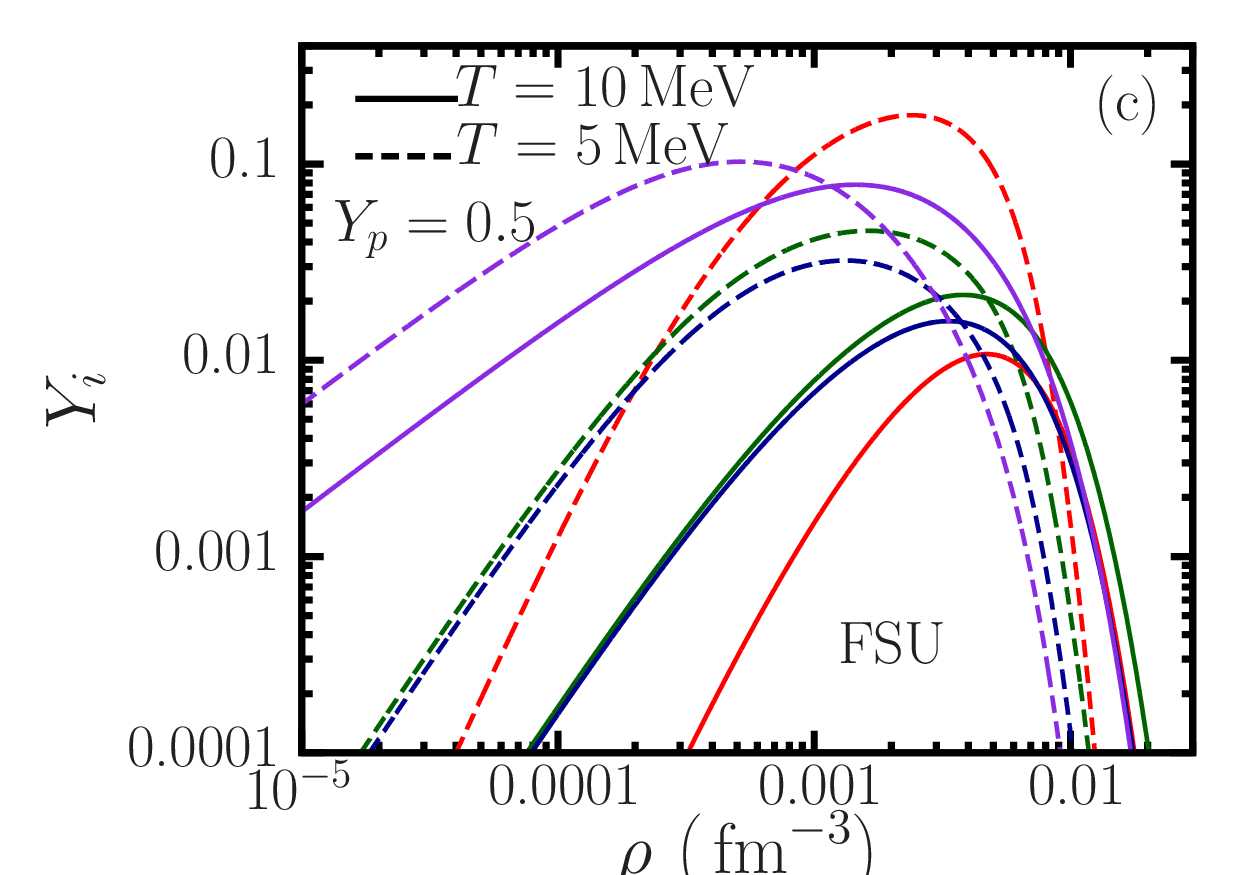}&
\includegraphics[width=0.45\linewidth,angle=0]{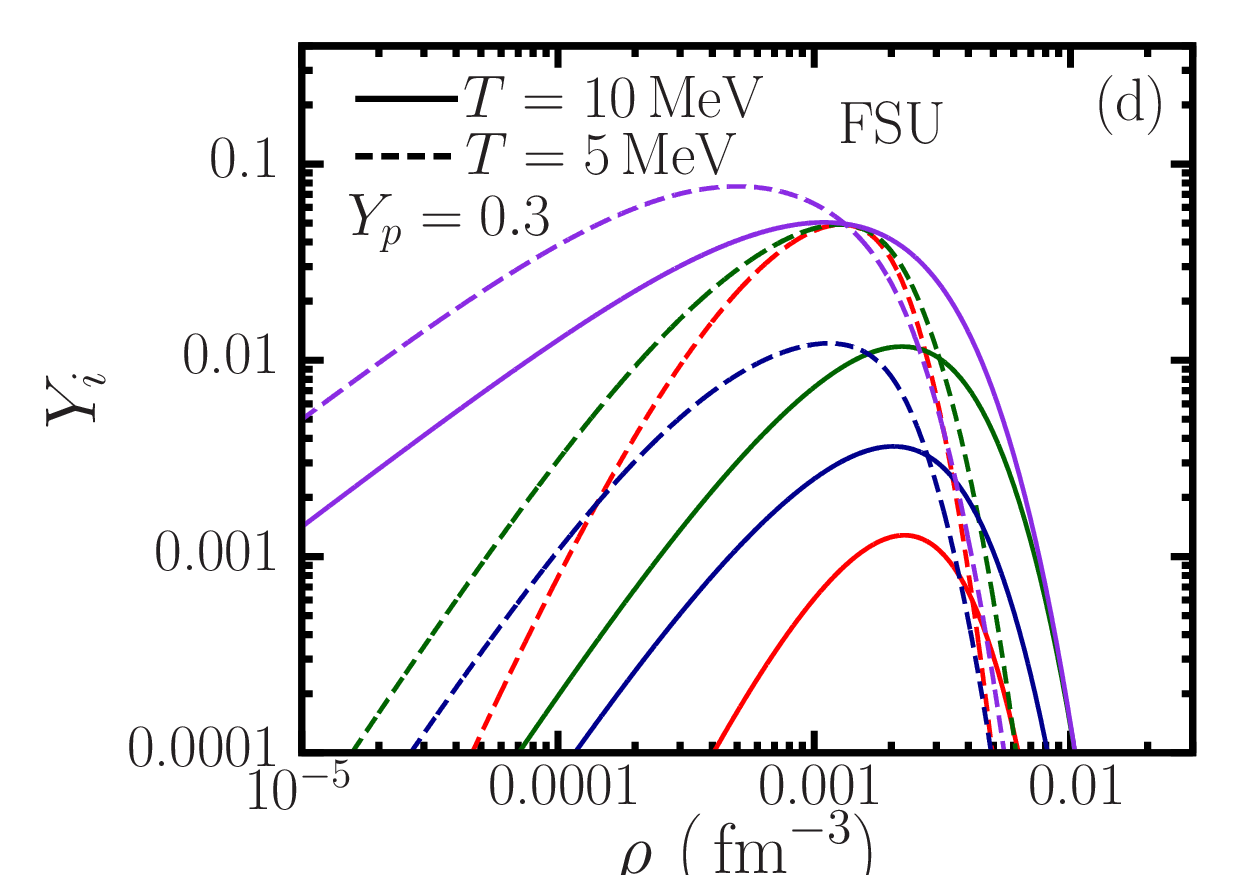}\\
\end{tabular}
\caption{(Color online) Fraction of light clusters in equilibrium with nuclear matter: (a) 
 for FSU with  $Y_p=0.5$ and $T=5$ MeV, and $g_{s}^i=0$ (dashed lines) and
 $g_{s}^i=3\, A_i/4\, g_s$ (full lines); (b)  for $g_{s}^i=3\, A_i/4\, g_s$ with
 $Y_p=0.5$ and $T=5$ MeV  and FSU (full lines) and NL3 (dashed  lines); (c)
 for FSU with   $g_{s}^i=3\, A_i/4\, g_s$ and $Y_p=0.5$, and $T=10$ MeV (full
 lines) and $T=5$ MeV  (dashed  lines);  (d)
 for FSU with   $g_{s}^i=3\, A_i/4\, g_s$ and $Y_p=0.3$, and $T=10$ MeV (full
 lines) and $T=5$ MeV  (dashed  lines).}
\label{fraction}
\end{figure*}

In the study by Typel {\it et al.} \cite{typel2010} the binding energy of the clusters in
the medium, 
\begin{equation}
B_i=A_i M^*-M^*_i
\label{bind}
\end{equation}
 where $M^*_i$ is the effective mass of cluster
$i$ defined in (\ref{Mstar}) and (\ref{mboson}), 
was
parametrized as a function of density and temperature. In Fig. \ref{mott}(a) we
compare the in-medium binding  energy of the $\alpha$-clusters as a
function of the density for several values of the couplings $g^i_{s}$. All
results were obtained at $T=5$ MeV but, in fact, they do not depend much
on the temperature. In the same figure we also include results obtained in
\cite{typel2010} at $T=5$ MeV, which turn out to be very close to the
calculated binding energies settings $g_{s}^i= 3\,
A_i/4 \, g_s$.
In order to reproduce the
results of  \cite{typel2010} we must consider temperature dependent
$\sigma$-cluster.
 We note, however, that choosing a temperature dependent parameter
will alter the usual thermodynamic relations between energy density and
pressure.  In Fig. \ref{mott}(b) we compare the in-medium binding energy of the
four light clusters obtained in our calculation setting
 $g_{s}^i= 3\, A_i/4 \, g_s$ with the results of  \cite{typel2010}: the best
accordance occurs for the $\alpha$-clusters. We may get a better accordance
for all the clusters if a different coupling fraction $g_{s}^i/g_s$ is taken
for each cluster.

 In Fig. \ref{exp} we compare the Mott densities, the densities at which the
binding energies defined in (\ref{bind}) vanish, at $T=5$  setting
$g_{s}^i= 3\, A_i/4 \, g_s$ (open triangles)
with the experimental Mott points determined in \cite{hagel2012} (full dots with errorbars) and the parametrization of the Mott point
given in \cite{typel2010} (dashed lines). Results in  \cite{hagel2012} are
obtained at different temperatures for the four light clusters, however, all
temperatures lie close to 5 MeV.  We conclude that choosing $g_{s}^i= 3\,
A_i/4 \, g_s$, and the corresponding $g_{v}^i$ and  $g_{\rho}^i$ couplings
defined in Table \ref{tab:fit},  at $T=5$ MeV in our model,  gives results that are in
reasonable agreement with the data of Ref. \cite{hagel2012} for the Mott density.



As a test of the parametrizations proposed for $g_s^i= 3A_i/4\, g_s$ in Table \ref{tab:fit},
we calculate the particle fraction at finite temperature and in chemical
equilibrium, $\mu_i=Z_i \mu_p+(A_i-Z_i) \mu_n$ with $i=\alpha, \, h,
\,t, \, d$ \cite{alphas,clusters}.
We first focus on the effect of $g_s^i$ and
 show in
Fig. \ref{fraction}(a) the fraction of light clusters at $T=5$ for
symmetric nuclear matter obtained for two values of the $\sigma$-cluster coupling
constant, 0 (dashed lines)  and $3A_i/4\, g_s$ (full lines). It is
seen that the particle fraction and dissolution density is sensitive to this
coupling: for a larger coupling, both  the fraction of particles and
 the dissolution density are larger. In Figs.  \ref{fraction}(b) --
 \ref{fraction}(d), we always set $g_s^i= 3A_i/4\, g_s$.
In Fig. \ref{fraction}(b) we compare two models, NL3 (dashed lines) and FSU
(full lines). It is seen that the results do not differ much except for the deuterons
that dissolve  at a larger density  and attain a larger fraction with NL3. 
At very low densities when the
interaction between nucleons and clusters is negligible both
models coincide; at densities just below the dissolution density,
where 
models differ the most,  the
differences are not very large. The effect of temperature and isopsin asymmetry is
shown in Fig. \ref{fraction}(c) and (d). In
Fig. \ref{fraction}(c) particle fractions in symmetric matter are compared for $T=10$ MeV (full lines) and  $T=5$ MeV (dashed lines). At $T=10$ MeV the deuteron
fraction is already the largest fraction and the $\alpha$ fraction the
smallest. In neutron rich matter, such as the one represented in
\ref{fraction}(d), the deuteron fraction is the largest and the tritons come
in second both at $T=5$ and $10$ MeV. The large amount of deuterons is in
agreement  with results of \cite{light2}, where composition of low-density
supernova matter composed of light nuclei was calculated within a
quasiparticle gas model, 
and \cite{light3}, where composition of the outer layers of a protoneutron
star was studied.

\section{Conclusions}\label{conclusions}
In the present work we have included light clusters in the EOS
of nuclear matter within the framework of relativistic mean field
models with constant coupling constants.
 Clusters are considered point-like particles that interact with the
nucleons through the exchange of mesons. The formation of clusters is
favorable at a quite low density; below 0.002 fm$^{-3}$, we expect
that it is a reasonable approximation to consider them as point like.

It was shown that the dissolution of clusters is mainly determined by the
isoscalar part of the EOS.
 Therefore,   the  $\rho$-cluster coupling constant was simply determined by the isospin
of the cluster. To fix the $\sigma$ and $\omega$ clusters couplings
 two constraints are required  for each cluster. In the present work we have
considered the dissolution
 density obtained in a
quantum statistical approach \cite{ropke} as a constraint to fix the
$\omega$-cluster coupling.  Recent experimental results for the in-medium binding energy
of light clusters \cite{hagel2012} allow the determination of the $\sigma$-cluster coupling
strength at $T\sim 5$ MeV.
 The virial expansion of the EOS
at low densities and finite temperature is another constraint that could be used
to fix the $\sigma$-meson coupling and that will be investigated.

We have applied the couplings proposed within the present to study
symmetric and asymmetric nuclear matter with light clusters in chemical
equilibrium at finite temperature and we have determined
the relative light cluster fractions at $T=5\unit{MeV}$ and $T=10\unit{MeV}$. It was shown that a
larger $\sigma$-cluster coupling gives rise to larger dissolution densities
and larger particle fractions.

The experimental determination of  Mott points at more temperatures than the
ones obtained in \cite{hagel2012} would allow the determination of the
temperature dependence. A comparison of the in-medium binding energies obtained within the present
model with the ones proposed in \cite{typel2010}, indicate that the last may
be reproduced if  temperature dependent meson-cluster couplings are obtained. 
 It was shown that deuterons and tritons are the clusters with larger
abundances in asymmetric matter above $T=5$ MeV.. The results do not dependent much on the RMF
interaction chosen.

{\bf Acknowledgments}: This work was partially supported by QREN/FEDER, the
Programme COMPETE,  under the project PTDC/FIS/113292/2009 and  by Compstar, an ESF Research
Networking Programme.

\end{document}